\hoffset = -10mm
\baselineskip=6mm
\vglue1cm \line {\vbox{\halign{\hfill#\hfill\cr Nice INLN 02/10 \cr}}
\hfill\vbox{\halign{\hfill#\hfill\cr October 2002\cr}}} \vglue4cm
\magnification=1150

\centerline { {\bf PROOF OF A MASS SINGULARITY FREE PROPERTY}}\bigskip\centerline{{\bf IN}}\bigskip
\centerline{{\bf HIGH TEMPERATURE $QCD$
}} 
 \bigskip\bigskip\bigskip \centerline {{${\rm{T.\ Grandou}} $ }}
\bigskip\medskip 
\centerline { {\it Institut Non Lin\'eaire de Nice UMR CNRS
6618; 1361, Route des Lucioles, 06560 Valbonne, France}}\centerline
{e-mail:grandou@inln.cnrs.fr}
\bigskip\bigskip\medskip\bigskip\bigskip \centerline{\bf ABSTRACT}
\bigskip\smallskip\medskip It is shown that three series of diagrams entering the calculation of some hot $QCD$
process, are mass (or collinear) singularity free, indeed. This generalizes a result which was recently
established
up to the third non trivial order of (thermal) Perturbation Theory.

\vfill  \eject

{\bf{I. Introduction}}\bigskip\bigskip\bigskip
During the past fourteen years, a
considerable amount of work has been devoted to the study of quantized
fields at high temperature and/or chemical potential$^1$ (high temperature, e.g, means higher than any
bare or renormalized mass involved in the theory). The inherent non perturbative character of thermal quantum field theories has been
recognized$^2$, and naive perturbation theories accordingly reorganized.
This is achieved by means of a {\it{Resummation Program}} ($RP$)$^3$, which, in the high temperature limit, must
be
used whenever one
is calculating processes involving Green's functions with soft external/internal
lines. The soft scale is defined to be on the order of $gT$ where $T$ is the
temperature and $g$ some relevant and small enough coupling constant, so as to {\it {decide}}, at least
formally$^4$,
of two separate hard
(on the order of $T$) and soft energy scales. The $RP$ is given by effective Feynman rules, consisting of
effective
field propagators and $n$-points proper vertices, all at a given leading order of approximation
which turns out to be $g^2T^2$, and is referred to as $HTL$ ({\it{Hard Thermal
Loops}}). While $HTL$ vertices are purely perturbative objects, effective
propagators are not, as they give rise to pole residues and dispersion laws that
do not admit perturbative series expansions in the coupling constant. In the course of practical
calculations, effective propagators are easily handled, relying on analyticity properties and Cauchy's
theorem.\smallskip
Endowed with most beautiful symmetries, the $RP$ is an effective theory that has led to a number
of satisfying results$^1$, but has also met two serious obstructions, emanating both from the
infrared ($I\!R$) sector$^{5,6}$.\medskip
Now, {\it{Resummations}} can in general be defined a number of consistent, still different ways. In this article,
we take advantage of a so called {\it{Perturbative Resummation}} scheme, hereafter denoted $P\!R$ for short,
previously introduced$^7$ in the context of the first obstruction$^5$, to address the problem of the soft real
photon emission rate of thermal
$QCD$$^6$. This problem is the following. When use is made of the {\it{Resummation Program}} to calculate the
soft real
photon emission rate, out of a Quark-Gluon plasma in thermal equilibrium, the answer comes out affected with a
{\it{collinear singularity}}. In the context of massless Quantum Field Theories, it may be worth recalling that
collinear singularities$^8$ manifest themselves as singularities of the angular integration, or equivalently, of
the
integration on the {\it{virtuality}}, $P^2=p_0^2-p^2$, and are thus also called {\it{mass singularities}}. Though
of a different nature, mass/collinear singularities are regrouped with singularities of the integration on three-
momenta $|{\vec{p}}|$, under the common spell of {\it{Infrared singularities}}.\smallskip
Several attempts to cure that $I\!R$ disease have been proposed ever since$^9$, which, though consistent at one
loop order, encounter further important difficulties when extended to higher number of loop calculations$^{10}$. Our present work is motivated by a recent study of the problem, projected out on a toy model, with the
conclusion that things come out very different according to the Resummation scheme in use, $RP$ or $P\!R$
$^{11}$.
Then, the
case of interest, that is hot $QCD$, has been studied in its first three non trivial perturbative orders :
Again, no collinear singularity did show up in a $P\!R$ resummation scheme. Moreover, both the questionable
nature of that singularity $^{12}$, and the very mechanism through which it comes about in an
$RP$ scheme, have been discovered thanks to an original comparison with the $P\!R$ scheme $^{13}$.\medskip
However encouraging, this first analysis of the $QCD$ case has only been performed up to three
loop
order. In order to see if a
$P\!R$ resummation scheme has any chance to avoid that serious problem, it is crucial to extend the
proofs of Ref.13 to any loop order, and this sets in, we think, the
strategical interest of the present analysis. By the same token, we note that the quite as much important
{\it{collinear enhancement}} problem$^{10}$, which comes out of the latter
if one tries to solve the
difficulty by
introducing a so called {\it{asymptotic thermal mass}}, $m_\infty$ $^{9}$, is circumvented also.\medskip 
The paper is organized as follows. Section 2 is a reminder of the collinear singularity problem under
consideration, while introducing elements and notations necessary to the next sections. In Section 3, topologies
involving only bare vertices, with $N(N')$ $HTL$-self energy insertions along the $P(P')$ internal fermionic
lines,
denoted by $(N,N';0)$, are investigated in details. To do so, the matter of Ref.7 is exploited so as to show
that any $(N,N';0)$ imaginary part is mass singularity free, or {\it{msf}} for short. The same property can be
established concerning the contributions attached to
$(N,N';1)$ topologies, with one $HTL$-vertex correction included, and this is Section 4. The results of both
previous sections are obtained on the basis of purely technical calculations, but it seems almost impossible to
proceed furhter along this line of approach :
For contributions of type $(N,N';2)$, involving two $HTL$-vertex corrections, functions at play are so
complicated
that they preclude any control of the ensuing integrals. Remarkably enough, though, Section 3 is able to
provide enough
information so as to initiate an efficient, global and conclusive approach by induction. On the other hand, used
right from the onset, an induction procedure does not appear to be, by itself,
fully
conclusive. This efficient articulation of Section 3 calculational approach, to an induction process is the
matter
of Section~5. All three series of diagrams are definitely shown to possess mass/collinear singularity free
imaginary parts, in the end.\smallskip Our
conclusions are gathered in Section 6, whereas an Appendix displays the technical complexities encountered by a
calculational approach to $(N,N';2)$-type diagrams.

\smallskip Throughout the article, we work in the $R/A$
real time formalism, which is based on Retarded/Advanced free field functions $^{14}$. Also, we
will be using the convention of upper case letters for quadrimomenta and lower case ones for their components,
writing, for example, $P=(p_0, {\vec p})$. Our conventions for labelling internal and external momenta can be
read
off Fig.1.

\bigskip\bigskip\bigskip
{\bf II. The soft real photon emission rate of hot QCD} \bigskip

It is convenient to work in the real time formalism with retarded/advanced $(R/A)$ field functions, where a
concise and elegant derivation of the famous collinear singularity can be achieved $^{15}$. The soft real photon
emission rate is essentially related to the imaginary part of the quantity ${\Pi_{RR}}^\mu_\mu(Q)$, trace of the
soft real photon
polarization tensor,
hereafter written as ${\Pi_{R}}(Q)$. At pure one loop order, this imaginary part is zero. However, when the
photon is soft, this result is incomplete and the {\it{Resummation Program}} must be used instead of bare thermal
Perturbation Theory. This amounts to keep the one loop diagram of ordinary Perturbation Theory, while replacing
bare vertices
and propagators by their {\it{HTL}}-dressed counterparts. In Feynman gauge, the resulting expression reads (with
$n_F$, the Fermi-Dirac statistical factor, defined without absolute value),
$$\displaylines{\Pi_R(Q)=i\int {{\rm d}^4P\over
(2\pi)^4}(1-2n_F(p_0))\ {\rm
disc}\ Tr\biggl\lbrace {}^\star S_R(P)\  {}^\star\Gamma_\mu(P_R,Q_R,-P'_A)\cr\hfill {}^\star S_R
(P')\ {}^\star\Gamma^\mu(P_R,Q_R,-P'_A)\biggr\rbrace\qquad(2.1)}$$The discontinuity is to be taken in the
energy variable $p_0$, by forming the
difference of $R$ and $A$-indiced $P$-dependent quantities. Within standart notations, the fermionic {\it{HTL}}
self energies, effective propagators and vertices are respectively given by
$$\Sigma_\alpha(P)=m^2\int {{\rm{d}}{\widehat K}\over 4\pi} {{{\rlap / \!\widehat K}} \over {{ \widehat
K}}\!\cdot\! P+i\epsilon_\alpha}\ ,\ \ \ \ m^2= C_F{g^2T^2\over 8}\ ,\ \ \ \ \alpha=R,A\eqno(2.2)$$

$${}^\star S_\alpha(P)={i\over {\rlap / \!P} - \Sigma_\alpha(P)} 
\eqno(2.3)$$

$${}^\star\Gamma_\mu(P_\alpha,Q_\beta,P'_\delta)=-ie\left(\gamma_\mu+\Gamma^{HTL}_\mu(P_\alpha,Q_\beta,P'_\delta)\right)\eqno(2.4)$$

$$\Gamma^{HTL}_\mu(P_\alpha,Q_\beta,P'_\delta)=m^2\int{{\rm d}{\widehat
K}\over 4\pi}{ {\widehat K}_\mu\  {{\rlap / \!\widehat K}}\over
({\widehat K}\!\cdot\! P+i\epsilon_\alpha)({\widehat
K}\!\cdot\! P'+i\epsilon_\delta)}\eqno(2.5)$$where ${\widehat
K}$ is the lightlike four vector $(1,{\widehat
k})$. In the sequel, it will reveal extremely useful to introduce a "self
energy four vector" (of course, not a genuine Lorentz-4-vector!), by writing, instead of standart expression
(2.2),
$$\Sigma_\alpha(P)\  \buildrel (def)\over {=}\ {\rlap /\!\Sigma}_\alpha(P)=\gamma\!\cdot\! \Sigma_\alpha(P)=
\gamma_\mu \ m^2\!\int {{\rm{d}}{\widehat K}\over 4\pi}  { \widehat K^\mu \over 
{\widehat {K}}\!\cdot\! P+i\epsilon_\alpha}\eqno(2.6)$$The $RP$ basic steps entering the soft real photon
emission
rate calculation of thermal $QCD$ are as follows. In view of
(2.1) and (2.4), one gets three types
of terms: A term
with two bare vertices $\Gamma_\mu^{(0)}$, two terms with one bare vertex $\Gamma_\mu^{(0)}$ and the other
$\Gamma^{HTL}_\mu$, and a term with two $HTL$ vertices $\Gamma_\mu^{HTL}$. In $QCD$, the first three terms pose
no
problem: Terms of second type entail a collinear singularity which, thanks to a $U(1)$-Ward identity, cancels out
with a similar singularity coming from the last term. A
residual
collinear singularity remains though, induced by the latter, and we therefore focus on that particular
contribution including two vertices $\Gamma^{HTL}_\mu$. One gets, 
$$\displaylines{\Pi_R(Q)=i\int {{\rm d}^4P\over
(2\pi)^4}(1-2n_F(p_0))\ {\rm
disc}\ Tr\biggl\lbrace {}^\star S_R(P){\Gamma^{HTL}}_\mu(P_R,Q_R,-P'_A)\cr\hfill {}^\star S_R
(P'){{\Gamma^{HTL}}}^\mu(P_R,Q_R,-P'_A)\biggr\rbrace\qquad(2.7)}$$ Then substituing the relevant $QCD$
expressions,
(2.2)-(2.5),
one can write, with the convention $\epsilon_R=+\epsilon$, 
$$\displaylines{ \Pi_R(Q)= -ie^2m^4\int {{\rm d}^4P\over
(2\pi)^4}(1-2n_F(p_0))\int{{\rm d}{\widehat
K}\over 4\pi}\int{{\rm d}{\widehat
K'}\over 4\pi}  \cr\hfill
{\rm
disc}\  {{\widehat
K}\!\cdot\!{\widehat
K'}\ \ Tr\left({}^\star S_R(P){{\rlap / \!\widehat K}}{}^\star S_R(P'){{\rlap / \!\widehat K'}}\right)\over
({\widehat K}\!\cdot\! P+i\epsilon)({\widehat
K}\!\cdot\! P'+i\epsilon)({\widehat K'}\!\cdot\! P+i\epsilon)({\widehat
K'}\!\cdot\! P'+i\epsilon)}\qquad(2.8)} $$
Because of the factor ${\widehat
K}\!\cdot\!{\widehat
K'}$ appearing in the numerator, there is no double pole but a simple collinear one at ${\widehat
K}={\widehat Q}$, whose residue just involes the $U(1)$ Ward identity alluded to above, that is,
$$m^2\int{{\rm d}{\widehat
K'}\over 4\pi}\ { [{\widehat Q}\!\cdot\!{\widehat K'}]\ {{\rlap / \!\widehat K'}}\over ({\widehat K'}\!\cdot\!
P+i\epsilon)({\widehat
K'}\!\cdot\! P'+i\epsilon)}={1\over q}\ [\ \rlap / \!\Sigma_R(P)-\rlap / \!\Sigma_R(P')\ ]\eqno(2.9)$$and yields
for
$\Pi_R(Q)$ the
expression
$$\displaylines{-i{e^2m^2\over q}\int {{\rm d}^4P\over (2\pi)^4}(1-2n_F(p_0)) 
\ {\rm disc}\ \  \int{{\rm d}{\widehat
K}\over 4\pi}{1\over ({\widehat K}\!\cdot\! P+i\epsilon)({\widehat
K}\!\cdot\! P'+i\epsilon)}\cr\hfill \times\  Tr\left( {}^\star S_R(P){{\rlap / \!\widehat Q}}{}^\star
S_R(P')[\ \rlap / \!\Sigma_R(P)-\rlap / \!\Sigma_R(P')\ ]\right)\qquad(2.10)}$$The discontinuity in $p_0$ can be
taken, and an appropriate choice of the integration
contour
in the $p_0$-complex plane allows to write 
$$\displaylines{\Pi_R(Q)=-2{e^2m^2\over q}\int {{\rm d}^4P\over (2\pi)^3}(1-2n_F(p_0))\int{{\rm d}{\widehat
K}\over 4\pi}{\delta({\widehat K}\!\cdot\! P)\over {\widehat
K}\!\cdot\! Q+i\epsilon}\cr\hfill
\times\  Tr\left( {}^\star S_A(P){{\rlap / \!\widehat Q}}{}^\star
S_R(P')\ [\ \rlap / \!\Sigma_A(P)-\rlap / \!\Sigma_R(P')\ ]\right)\qquad(2.11)}$$where a factor of $2$ accounts
for
the two possibilities
${\widehat K}={\widehat Q}$ and ${\widehat
K'}={\widehat Q}$, and where the relation $P'=P+Q$ has been used. The angular integration develops a collinear
singularity at
${\widehat K}={\widehat Q}$, and is responsible for that singular part of
$\Pi_R(Q)$ which can be expressed as 
$$\displaylines {-{2{e^2m^2\over q}}\left( \int{{\rm d}{\widehat
K}\over 4\pi}{1\over Q\!\cdot\! {\widehat
K}+i\epsilon}\right)
\int {{\rm d}^4P\over
(2\pi)^3}\ \delta(P\!\cdot\!{\widehat Q})\ (1-2n_F(p_0))
\cr\hfill \times\  
Tr\left({}^\star S_A(P){{\rlap / \!\widehat Q}}{}^\star S_R(P')\ [\ \rlap / \!\Sigma_A(P)-\rlap / \!\Sigma_R
(P')\ ]\right)\qquad(2.12) }$$The two terms involving one bare vertex $\gamma_\mu$ and a one loop
$HTL$ correction $\Gamma^{HTL}_\mu$, entail a similar singularity which, when combined with (2.12), leave
uncancelled the $\Pi_R(Q)$ singular contribution
$$\displaylines {-{2i{e^2m^2\over q^2}}\left( \int{{\rm d}{\widehat
K}\over 4\pi}{1\over {\widehat
Q}\!\cdot\! {\widehat
K}+i\epsilon}\right)
\int {{\rm d}^4P\over
(2\pi)^3}\ \delta(P\!\cdot\!{\widehat Q})\ (1-2n_F(p_0))
\cr\hfill \times\  
[Tr\left({}^\star S_A(P){{\rlap / \!\widehat Q}}\right)-Tr\left({}^\star S_R(P'){{\rlap / \!\widehat
Q}}\right)]\qquad(2.13) }$$It is this result which, in the literature$^6$ is most usually written in the form
$$ {C^{st}\over\varepsilon}\int
{{d^4P}\over (2\pi)^4}\delta({\widehat{Q}}\!\cdot\! P)\ (1-2n_F(p_0))\sum_{s=\pm
1,V=P,P'}\pi(1-s{v_0\over v})\beta_s(V) \eqno(2.14)$$where the overall $1/\varepsilon$ results of a dimensionally
regularized evaluation of the factored out angular integration of (2.13), and where $\beta_s(V)$ is related to
the effective fermionic propagator usual parametrization$^{16}$,
$${}^\star S_{R,A}(P)=i\sum_{s=\pm 1}{{{\rlap / \!\widehat
P_s}}\over D_{R,A}^s(p_0,{\vec{p}})}\eqno(2.15)$$with ${\widehat{P_s}}=(1,s{\widehat{p}})$, the
label
$s$ referring to the two dressed fermion propagating  modes. Then one has,
$${1\over D^s_{R,A}(V)}=\alpha_s(V)\mp i\pi\beta_s(V)\eqno(2.16)$$

\bigskip\bigskip\bigskip

{\bf{III. Self Energy Diagrams, of type $(N,N';0)$}}\bigskip
The imaginary part of a general term of type $(N,N';0)$, depicted in Fig.1, can be
written 
$$2e^2\int
{{\rm d}^4P\over (2\pi)^4}(1-2n_F(p_0)) Tr\ \rlap /\!P {\rm{disc}}_P\left(\left(\rlap /\!\Sigma_R(P)\rlap /\!
P\right)^{N}\over (P^2)^{N+1}\right)\rlap /\!P' {\rm{disc}}_{P'}\left(\left(\rlap /\!\Sigma_R(P')\rlap /\!
P'\right)^{N'}\over (P'^2)^{N'+1}\right)\eqno(3.1)$$where the "Self Energy
four-vector" (2.2), has components, 
$$\Sigma_\alpha^0(P)={m^2\over 2p}\ln({p_0+p\over p_0-p})\ ,\ \ \ \ \Sigma_\alpha^i(P)=({{\vec p}^i\over p}\equiv
{\widehat p}^i) {m^2\over p}Q_1({p_0\over p})\eqno(3.2)$$with $Q_1$ standing for the Legendre function of
the second kind $$Q_1(x)=xQ_0(x)-1\ ,\ \ \ \ \ \ Q_0(x)={1\over 2}\ln({x+1\over x-1})\eqno(3.3)$$The label
$\alpha=\{R,A\}$
denoting one of the two {\it{Retarded}} or {\it{Advanced}} specifications of the real time formalism being used,
in the right hand sides of (3.2) these specifications are encoded in the logarithmic determinations.
\bigskip

It is elementary to prove that one has $$\left(\rlap /\!\Sigma_R\rlap /\!
P\right)^{N}=a_N\ \rlap /\!\Sigma_R\rlap /\!
P\ +b_N\ I\!\!I_4\eqno(3.4)$$where $I\!\!I_4$ is the $4\times 4$
identity matrix, and the coefficients $a_N$ and $b_N$
are polynomials in the variables $P\!\cdot\!\Sigma_R(P)=m^2$ and $-P^2\Sigma_R^2$ whose formation laws can be
found
to be given by  $$a_N=(m^2)^{N-1}\sum_{j=0}^{j_{M}(N)}C_N^{2k+1}\left(1-{P^2\Sigma^2\over
m^4}\right)^k\eqno(3.5)$$
$$b_N=(m^2)^N(-{P^2\Sigma^2\over
m^4})\sum_{j=0}^{j_{M}(N-1)}C_{N-1}^{2k+1}\left(1-{P^2\Sigma^2\over
m^4}\right)^k\eqno(3.6)$$The $C_N^{2k+1}$ are the binomial coefficients, and $j_M$, the maximal value of $j$ can
be expressed as $$j_{M}(N)={\left(N-1-\Theta\left((-1)^N\right)\right)\over 2}\eqno(3.7)$$ where $\Theta(x)$ is
the usual
Heaviside
step function.\bigskip
Because of the decomposition law (3.4), four types of trace factors are found, that are
$$4P\!\cdot\! P'\eqno(3.8)$$
$$8m^2 P\!\cdot\! P'-4P^2P'\!\cdot\!\Sigma_R(P)\ +\ (P\leftrightarrow P')\eqno(3.9)$$
$$(3.9)\left(P\leftrightarrow P'\right)\eqno(3.9')$$
$$(2m^2)^24P\!\cdot\! P'-\left(8m^2P'^2P\!\cdot\!\Sigma_R(P')+(P\leftrightarrow P')\right)+4P^2P'^2 
\Sigma_R(P)\!\cdot\!\Sigma_R(P')\eqno(3.10)$$\medskip\noindent We will therefore begin with proving that
integrals
of the generic
type $$\int
{{\rm d}^4P\over (2\pi)^4}(1-2n_F(p_0))\ {\rm{disc}}_P\left({(-P^2\Sigma^2_R(P))^{n}\over (P^2+i\epsilon
p_0)^{N+1}}\right){\rm{disc}}_{P'}\left({(-{P'}^2\Sigma^2_R(P'))^{n'}\over ({P'}^2+i\epsilon
p'_0)^{N'+1}}\right)\eqno(3.11)$$are mass singularity free, or {\it{msf}}, for short. Then, since all of
the trace factors (3.8)-(3.10) come into play as multiplicative functions of the integrands appearing in (3.11),
we
will check
that they leave unaltered its {\it{msf}} character.\smallskip
Integrals of generic type (3.11) : With $y={\widehat q}\!\cdot\!{\widehat p}$, where ${\widehat
q}$ and ${\widehat p}$ are the unit three-vectors in the directions of ${\vec{q}}$ and ${\vec{p}}$ respectively,
integration on $y$
can be traded for an integration on the virtuality $P'^2=-x'p'^2(y)$ by writing,
$$\int^1_{{P^2+2qp_0\over 2qp}} {\rm{d}}y={(p_0+q)^2\over 2qp}\int^{1-{(p_0+q)^2\over (p+q)^2}}_0
{{\rm{d}}x'\over (1-x')^2}\eqno(3.12)$$where the restrictions on $y$ and $x'$ come from the $\Theta(-P'^2)$
support of the distribution to be folded in (3.11). Now, particular to the thermal context$^{17}$, so called
{\it{Lebesgue non-integrable}} mass (and/or $I\!R$) singularities do arise, which cannot be
taken care of by means of a standart dimensional regularization procedure, and require that an extra $I\!R$
regulator be introduced$^{18}$. This is achieved by proceeding to the following replacement
$${1\over ({P'}^2+i\epsilon
p'_0)^{N'+1}}\longmapsto {1\over ({P'}^2-\mu^2+i\epsilon
p'_0)^{N'+1}} \eqno(3.13)$$that is also,
$$\delta^{(N')}(P'^2)\longmapsto \delta^{(N')}(P'^2-\mu^2)\ ,\ \ \ \ \ \ \
\ \ \ {{\bf P}\over (P'^2)^{N'+1}}\longmapsto {{\bf P}\over
(P'^2-\mu^2)^{N'+1}}\eqno(3.14)$$
where, as shown in Ref.7, Appendix $B$, the auxiliary $I\!R$
regulator $\mu^2$, is choosen a small, negative parameter, to be taken to zero in the end. Gathering pieces,
integration on $y$ can eventually be written as
$$\displaylines{{(m^4)^{n'}\over \left(-\mu^2+(p_0+q)^2\right)^{N'+1}}\ (-1)^{N'+1}{(p_0+q)^2\over
2qp}\int^{1-{(p_0+q)^2\over
(p+q)^2}}_0 {{\rm{d}}x'\over (1-x')^2}\ \ (1-x')^{N'+1}\ (x')^{n'}\cr\hfill
\biggl\lbrace {{\bf{P}}\over (x'-\lambda')^{N'+1}}\ Im\left(-1+{x'\over
4}[\epsilon(p'_0)\ln X']^2+{\sqrt{1-x'}}\ [\epsilon(p'_0)\ln X']\right)^{n'}\cr\hfill
+\pi\epsilon(p'_0){(-1)^{N'}\over N'!}\delta^{(N')}(x'-\lambda')\ Re\left(-1+{x'\over
4}[\epsilon(p'_0)\ln X']^2+{\sqrt{1-x'}}\ [\epsilon(p'_0)\ln X']\right)^{n'}\biggr\rbrace\qquad(3.15)}$$where
we have defined  $$\lambda'\ \buildrel def\over {=}\  {-\mu^2\over -\mu^2+(p_0+q)^2}\ ,\ \ \ \ \ \ X'\ \buildrel
def\over
 {=}\ {\epsilon(p'_0){\sqrt{1-x'}}+1\over  \epsilon(p'_0){\sqrt{1-x'}}-1}\eqno(3.16)$$and where $\epsilon(p'_0)$
is
 the distribution "sign of $p'_0$".
 The    remaining  two integrations
are
 on  $p=|{\vec{p}}|$, and $p_0$, and the latter can be translated into an integration on the virtuality variable
 $x=-P^2/p^2$.\bigskip Now, if we consider the integral,
 $$\int_{-p}^{+p}{\rm{d}}p_0\ (1-2n_F(p_0))\ {\rm{disc}}_P\left({(-P^2\Sigma^2_R(P))^{n}\over (P^2+i\epsilon
p_0)^{N+1}}\right)\eqno(3.17)$$which enters (3.11) as a building block:
$$(3.11)=\int{p^2{\rm{d}}p\over (2\pi)^3}\int_{-p}^{+p}{\rm{d}}p_0 (1-2n_F(p_0))
{\rm{disc}}_P\left({(-P^2\Sigma^2_R(P))^{n}\over (P^2+i\epsilon p_0)^{N+1}}\right)\times
(3.15)\eqno(3.18)$$
we get for (3.17) the expression,       $$\displaylines{{1\over 2}\ p\ ({-1\over
p^2})^{N+1}(m^4)^n\sum_{\epsilon(p_0)=\pm 1}\int_0^1{{\rm{d}}x\over  {\sqrt{1-x}}}\     
(1-2n_F(\epsilon(p_0)p{\sqrt{1-x}}))\  x^n\biggl\lbrace 
{{\bf{P}}\over
(x-\lambda)^{N+1}} 
\cr\hfill\times Im\left(-1+{x\over 4}[\epsilon(p_0)\ln X]^2+{\sqrt{1-x}}[\epsilon(p_0)\ln  X]\right)^{n}\cr\hfill
 +\pi\epsilon(p_0){(-1)^{N}\over N!}\delta^{(N)}(x-\lambda)\cr\hfill  \times Re\left(-1+{x\over
4}[\epsilon(p_0)\ln
 X]^2+{\sqrt{1-x}}[\epsilon(p_0)\ln X]\right)^{n}\biggr\rbrace\qquad(3.19)}$$with the definitions,
 $$\lambda \ \buildrel def\over {=}\  {-\mu^2\over p^2}\ ,\ \ \ \ \ \ X\ \buildrel def\over {=}\ 
{\epsilon(p_0){\sqrt{1-x}}+1\over \epsilon(p_0){\sqrt{1-x}}-1}\eqno(3.20)$$Eventually, an integration on  $p$
must
be
performed, which can symbolically be written as
$$\int_{p_m}^{p^\star}{p^2{\rm{d}}p\over (2\pi)^3} \ \times\ G\left(p,q,m\right)
\eqno(3.21) $$ An upper bound of integration on $p$ is introduced so as to avoid the
hard region $p={\cal{O}}(T)$ (a customary choice consists in taking ${p^\star}$ on the order of an
intermediate scale, say, on the order of ${\sqrt{g}}T$), whereas the lower boundary, $p_m$, not relevant to our
concern here, will be discussed elsewhere$^{19}$. Note that in (3.15) and (3.19), we have written between
brackets expressions of the form
$[\epsilon(p_0)\ln  X]$. This is because, irrespective of
the
sign of $p_0$, these expressions can be
written with the help of a most efficient representation $^{7,11,13}$
$$\epsilon(p_0) \ln X(p_0)=\epsilon(-p_0) \ln X(-p_0)=\lim_{\varepsilon=0}\
{1\over \varepsilon}\ \left(\ 1-{x^\varepsilon e^{i\pi\varepsilon}\over (1+{\sqrt{1-x}})^{2\varepsilon}}\
\right)\eqno(3.22)$$ Thanks to the $x^{\varepsilon}$ factor, this
representation is able to provide mass/collinear singularities with the same regularization as a
dimensional one would operate, while being far simpler. It is also endowed with interesting regularity
properties, since, in particular, the limit $\varepsilon=0$, commutes with both the sum over $N$ and integral on
$p_0$$^{7}$.
\smallskip
Using (3.22), one obtains an expansion $$\displaylines{\left(-1+{x\over 4}[\epsilon(p_0)\ln X]^2
+{\sqrt{1-x}}[\epsilon(p_0)\ln X]\right)^{n}=\sum_{i=0}^nC_n^i(-1)^i({x\over
4})^{n-i}\sum_{k=0}^iC_i^k(-1)^k{\sqrt{1-x}}^k\cr\hfill\times\  {1\over
\varepsilon^{2(n-i)+k}}\sum_{m=0}^{2(n-i)+k} C_{2(n-i)+k}^m\ (-1)^m{e^{i\pi m\varepsilon}x^{m\varepsilon}\over
(1+{\sqrt{1-x}})^{2
m\varepsilon}}\qquad(3.23)} $$which can be put back into (3.19). Introducing the following family of functions
$${\cal{F}}_{k-1,m}(m\varepsilon,x)={
{\sqrt{1-x}}^{k-1} \over
(1+{\sqrt{1-x}})^{2m\varepsilon}}(1-2n_F(\epsilon(p_0)p{\sqrt{1-x}}))\eqno(3.24)$$we note that it is
convenient
to proceed as
for the moving fermion damping rate problem$^{7}$, keeping the
leading order term of the statistical weight high temperature expansion
$${\cal{F}}_{k-1,m}(m\varepsilon,x)={p\epsilon(p_0)\over 2T}{
{\sqrt{1-x}}^{k} \over
(1+{\sqrt{1-x}})^{2m\varepsilon}}\left(1+{\cal{O}}(g^2)\right)\ \buildrel def\over {=}\ {p\epsilon(p_0)\over 2T}\
{F}_{km}(m\varepsilon,x)\ \left
(1+{\cal{O}}(g^2)\right)\eqno(3.25)$$Though no way mandatory (the same results being obtained otherwise), this
simplification is consistent with the leading order
calculation we are concerned with, preserves the correct parity in $p_0$, and allows to recognize in
$F_{km}$ the same expression as defined in Ref.7, Eq.(7.2). Whereof we know that (3.19) is rigorously {\it
{integrable}} and {\it {non-integrable}} mass
singularity free: The expression (3.19) can effectively be written as
$$\displaylines{{1\over 4T}\ p^2\ ({-1\over
p^2})^{N+1}(m^4)^n\sum_{\epsilon(p_0)=\pm 1}\sum_{i=0}^nC_n^i(-1)^i({1\over
4})^{n-i}\sum_{k=0}^iC_i^k(-1)^k\cr\hfill {1\over
\varepsilon^{2(n-i)+k}}\sum_{m=0}^{2(n-i)+k}C^m_{2(n-i)+k}{(-1)^m
 }\int_0^1{\rm{d}}x\ \ \biggl\lbrace {\bf{P}}{{\cal{I}}m(e^{i\pi
m\varepsilon})\over
(x-\lambda)^{N+1}}\cr\hfill +{\cal{R}}e (e^{i\pi m\varepsilon})\ \pi\epsilon(p_0){(-1)^{N}\over
N!}\delta^{(N)}(x-\lambda)\biggr\rbrace x^{2n-i+m\varepsilon}{F_{km}(m\varepsilon,x)}\qquad(3.26)}$$which is
Eq.(D.3) of Ref.7. Mass singularities of the {\it{non-integrable}} type, ${\cal{O}}(1/\lambda)^k$, cancel out
$$\displaylines{\int_0^1{\rm{d}}x\ \ \biggl\lbrace {\bf{P}}{{\cal{I}}m(e^{i\pi m\varepsilon})\over
(x-\lambda)^{N+1}}\cr\hfill +{\cal{R}}e (e^{i\pi m\varepsilon})\ \pi\epsilon(p_0){(-1)^{N}\over
N!}\delta^{(N)}(x-\lambda)\biggr\rbrace x^{2n-i+l+m\varepsilon}={\sin(\pi m\varepsilon)\ \over
2n-i+l-N+m\varepsilon}+{\cal{O}}(\lambda)\qquad(3.27)}$$whereas {\it{integrable}} mass singularities obey arithmetical cancellation
patterns thanks to the identities
$$\bigl\lbrace {\varepsilon^p\over
\varepsilon^j} \bigr\rbrace\times\sum_{m=0}^jC_j^m(-1)^m\ m^p=0\ ,\ \
\ \ 1\leq p\leq j-1\eqno(3.28)$$
$$\sum_{m=0}^jC_j^m(-1)^m\
m^j=(-1)^jj!\eqno(3.29)$$It can even be shown (Appendix D of Ref.7) that (3.26) defines a mapping of $C\times C$
into $C$ which is analytic for $(\varepsilon,\lambda)$ choosen in the product of discs $D(0,{1\over 2N})\times
D(0,{1\over 2})$. The limit $\varepsilon=0,\lambda=0$ therefore exists and is independent of the sequence along
which it is taken.

\bigskip\smallskip The entirely new feature is
of course that the integrand appearing in (3.19), gets
supplied, now,
with the extra function (3.15). Considered as a function of $x$, the properties of (3.15) are therefore crucial
in
order to address the ensuing behaviour
of generic type (3.11) integrals, and this is what we now turn to examine in the particular case
of positive energies, $p_0>0$, for the sake of a simpler illustration. \bigskip As made obvious by inspection,
(3.15) is essentially
relevant of the same structure as displayed by
(3.19): Up to an overall multiplicative function of $p_0$,
$${(m^4)^{n'}\over \left(-\mu^2+(p_0+q)^2\right)^{N'+1}}\ {(p_0+q)^2\over
2qp}\eqno(3.30)$$ the difference is entirely
in the integration range $$0\leq x'\leq x'_M(x)\ \buildrel def\over {=}\ 
1-{\left(p_0(x)+q\right)^2\over(p+q)^2}\eqno(3.31)$$ instead
of $0\leq x\leq
1$. Using the
representation (3.22) for the expression
$[\epsilon(p'_0)\ln X']$, and the binomial expansion (3.23), the same
functions as in (3.25) can
effectively be identified, with accordingly, the same properties $$F_{2(N'-1)+k',\ m'}\ (m'\varepsilon',x')={
{\sqrt{1-x'}}^{2(N'-1)+k'} \over
(1+{\sqrt{1-x'}})^{2m'\varepsilon'}}\eqno(3.32)$$In the limit of $\lambda'=0$, we learn out of Ref.7, that 
{\it{ Lebesgue
non-integrable}} mass singularities cancel out: Up to the overall multiplicative factor (3.30), one is left for
the
full
expression (3.15), with an expression which still displays a finite series of {\it{Lebesgue
integrable}}
mass singularities:
$$\displaylines{\sum_{l'=0}^\infty\ {1\over
l'!}\ \sum_{i'=0}^{n'}C_{n'}^{i'}(-1)^{i'}
\ \sum_{k '=0}^{i'}C_{i'}^{k'}(-1)^{k'} \left(g(p_0){(q+p)^2\over -P^2}\right)^{N'-(2n'-i'+l')}\cr\hfill\times
{1\over \varepsilon'^{2(n'-i')+k'}}\sum_{m'=0}^{m'=2(n'-i')+k'}C_{2(n'-i')+k'}^{m'}\ (-1)^{m'} {\sin(\pi
m'\varepsilon')\over N'-(2n'-i'+l')-m'\varepsilon'}\cr\hfill \times \left(g(p_0){(q+p)^2\over
-P^2}\right)^{-m'\varepsilon'}
F^{(l')}_{2(N'-1)+k',\ m'}\ (m'\varepsilon',0)\qquad(3.33)}$$where we have defined the function,  $$g(p_0)\
\buildrel
def\over {=}\  {p+p_0\over p+p_0+2q}=1-{2q\over 2q+p}\ (1+{p\over 2q+p}{\sqrt{1-x}})^{-1}\eqno(3.34)$$Type
${\cal{O}}({\varepsilon'}^{-r})$-mass singularities are thus controlled by the finite sum,
$$\displaylines{{1\over \varepsilon'^{2(n'-i')+k'}}\sum_{m'=0}^{m'=2(n'-i')+k'}C_{2(n'-i')+k'}^{m'}\ (-1)^{m'}
\  
\left({x\over g(p_0)}\right)^{m'\varepsilon'}\cr\hfill \times\ {\sin(\pi
m'\varepsilon')\over N'-(2n'-i'+l')-m'\varepsilon'}\times \left(({p\over p+q
})^2\right)^{m'\varepsilon'}\times\  F^{(l')}_{2(N'-1)+k',\
m'}\ (m'\varepsilon',0)
\qquad(3.35)}$$ where the last term appearing in the right hand side of (3.35), stands for the ${l'}^{th}$-order
derivative of the function (3.32), taken
at $x'=0$. This means that, in deriving (3.35), we have interchanged the sum on $l'$, in the Taylor
expansion
of (3.32),
$$F_{2(N'-1)+k',\ m'}\ (m'\varepsilon',x')=\sum_{l'=0}^\infty {(x')^{l'}\over
l'!}\ F^{(l')}_{2(N'-1)+k',\ m'}\
(m'\varepsilon',0)\eqno(3.36)$$with the integration on $x'$. Such a permutation is proven to be licit in
Ref.7, Appendix $C$ (note that in the present situation, this permutation is the more licit, as the integration
range (3.31) lies within the unit convergence radius of the series expansion for the
functions $F_{2(N'-1)+k',\ m'}\ (m'\varepsilon',x')$).
Likewise, it is demonstrated (Eqs. (C.7)-(C.9), (C.12) of Ref.7) that each of the coefficients
$F^{(l')}_{2(N'-1)+k',\ m'}\ (m'\varepsilon',0)$
admits a Taylor series expansion in the parameter $m'\varepsilon'$. Now, whatever $N'$ and
$2n'-i'+l'$,
the same property holds clearly true, for any of the other three factors of (3.35) which, with
$F^{(l')}_{2(N'-1)+k',\ m'}
(m'\varepsilon',0)$, enter the sum over
$m'$.\smallskip Then, forming the Cauchy's product of their $m'\varepsilon'$-Taylor series expansions,
and relying on the set of arithmetical identities (3.28) and (3.29), we conclude that the $\varepsilon'=0$ -limit
of (3.35) is
{\it{msf}}, and reduces to a polynomial of degree $2(n'-i')+k'$ in the variable $\ln
\left(x/g(p_0)\right)$,$$\lim_{\varepsilon'=0}\ (3.35)
=(-{1})^{2(n'-i')+k'}  \  \sum_{j'=0}^{2(n'-i')+k'}\ C_{2(n'-i')+k'}^{j'}\ 
H^{(2(n'-i')+k'-j')}(0)\ \ln^{j'} ({x\over g(p_0)})\eqno(3.37)$$where $H^{(2(n'-i')+k'-j')}(0)$, a
pure (real) number, is a shorthand notation for the derivative of order $(2(n'-i')+k'-j')$, taken at
$m'\varepsilon'=0$, of the
product $${\sin(\pi
m'\varepsilon')\over N'-(2n'-i'+l')-m'\varepsilon'}\ \times\  \left(({p\over
p+q})^2\right)^{m'\varepsilon'}\times\  F^{(l')}_{2(N'-1)+k',\
m'}\ (m'\varepsilon',0)\eqno(3.38)$$

\bigskip\noindent
Gathering all pieces, the whole expression (3.15) can eventually be written as
$$\displaylines{-{(m^4)^{n'}\over \left(-\mu^2+(p_0+q)^2\right)^{N'+1}}\ {(p_0+q)^2\over 2qp}\cr\hfill\times\ 
\sum_{l'=0}^\infty\ {(-1)^{l'}\over l'!}\ \sum_{i'=0}^{n'}C_{n'}^{i'}\ \left(g(p_0){(q+p)^2\over
+P^2}\right)^{N'-(2n'-i'+l')}
   \cr\hfill
  \times\ \sum_{k '=0}^{i'}C_{i'}^{k'}\  \sum_{j'=0}^{2(n'-i')+k'}
H^{(2(n'-i')+k'-j')}(0)\ \ln^{j'} ({x\over g(p_0)})\qquad(3.39)}$$Getting back to (3.18), one is now in a
position
so as to estimate the incidence on (3.19) of any of the extra $x$-dependences which are introduced by
(3.39).\smallskip For positive (as well as negative) energies, the auxiliary $I\!R$ regulator
$-\mu^2$ can safely be taken to zero in the prefactor of (3.39), and the latter expanded as $${-(m^4)^{n'}\over
\left(-\mu^2+(p_0+q)^2\right)^{N'+1}}\ {(p_0+q)^2\over 2qp}=-{(m^4)^{n'}\over
2pq}\sum_{r=0}^\infty {\cal{C}}_k(N',p,q)\ {\sqrt{1-x}}^r\eqno(3.40)$$where it easy to check that the existence
of
(3.40) does not depend on the relative magnitude of $p$ and $q$, contrarily, of course, to the explicit form of
the
coefficients ${\cal{C}}_k(N',p,q)$. The same property is obviously shared by the function
$$g(p_0)^{N'-(2n'-i'+l')}=\left(1-({2q\over 2q+p})\ {1\over 1+{p\over
2q+p}{\sqrt{1-x}}}\right)^{N'-(2n'-i'+l')}\eqno(3.41)$$ Eventually, such is also the case of factors like
$$\left(\ln{1\over
g(p_0)}\right)^{r'}=(-1)^{r'}\ln^{r'}\left(1-({2q\over 2q+p})\ {1\over 1+{p\over
2q+p}{\sqrt{1-x}}}\right)\eqno(3.42)$$The product of terms (3.40)-(3.42) can therefore be written as a
series in the variable ${\sqrt{1-x}}$, whose general term, no matters how complicated, just
redefines the integer power $k$ of the function $F_{km}(m\varepsilon,x)$ introduced in (3.25). The properties of
(3.25) are
thus left the same, and the extra $x$-dependences introduced through (3.40)-(3.42) preserve the {\it{msf}}
character of (3.19).\medskip 
The extra factors of (3.39),
$$({(q+p)^2\over P^2})^{N'-(2n'-i'+l')}\eqno(3.43)$$redefine the power $N+1$ of the scalar propagator appearing
in (3.11), according to the replacement $${1\over (P^2-\mu^2+i\epsilon p_0)^{N+1}}\ \longmapsto\  {1\over
(P^2-\mu^2+i\epsilon p_0)^{N+1+N'-(2n'-i'+l')}}$$This splits into the
distributions
$$\delta^{\left(N+N'-(2n'-i'+l')\right)}(P^2-\mu^2)\ ,\ \ \ \ \ \ \
\ \ \ {{\bf P}\over
(P^2-\mu^2)^{N+1+N'-(2n'-i'+l')}}$$which, with respect to the previous power of $N+1$, require extra
differentiability of the $x$-dependences they act upon. Now, this condition is clearly met thanks both to a full
identification of the new $x$-dependences brought about by (3.39), and to
the introduction of the auxiliary $I\!R$
regulator $\lambda=-\mu^2/p^2$ of (3.20). Since, at $N\geq 2n-i+l+1$ (which is just the condition for the
occurence
of
mass singularities), the overall compensation of mass
singularities does not depend on the relative magnitude of the integers $N+1$ and $2n-i+l$, in mass singularity
compensation patterns of the generic type$^{7}$,
$$\displaylines{\lim_{\varepsilon,\lambda=0}\ {1\over
\varepsilon^{2(n-i)+k}}\sum_{m=0}^{2(n-i)+k}C^m_{2(n-i)+k}{(-1)^m\
{F_{km}^{(l)}(m\varepsilon,0)} }\int_0^1{\rm{d}}x\ \ \biggl\lbrace {\bf{P}}{{\cal{I}}m(e^{i\pi
m\varepsilon})\over
(x-\lambda)^{N+1}}\cr\hfill +{\cal{R}}e (e^{i\pi m\varepsilon})\ \pi\epsilon(p_0){(-1)^{N}\over
N!}\delta^{(N)}(x-\lambda)\biggr\rbrace x^{2n-i+l+m\varepsilon}\ =\ {\cal{O}}(1)\qquad(3.44)}$$we deduce that the
same
mass singularity
compensations hold true of (3.44), with $N+1+N'-(2n'-i'+l')$ replacing $N+1$, and that extra factors of type
(3.43)
are {\it{msf}}-preserving.
\medskip
Eventually, the last extra $x$-dependences introduced into (3.18) by (3.39), are the functions
$$\ln^{r'} x\ ,\ \ \ r'\in I\!\!N\ ,\ \ \ 0\leq r'\leq 2(n'-i')+k'\eqno(3.45)$$Previous patterns (3.44) are now
taken
to the form $$\displaylines{{1\over
\varepsilon^{2(n-i)+k}}\sum_{m=0}^{2(n-i)+k}C^m_{2(n-i)+k}{(-1)^m\
{F_{km}^{(l)}(m\varepsilon,0)} }\int_0^1{\rm{d}}x\ \ \biggl\lbrace {\bf{P}}{{\cal{I}}m(e^{i\pi
m\varepsilon})\over
(x-\lambda)^{N+1}}\cr\hfill +{\cal{R}}e (e^{i\pi m\varepsilon})\ \pi\epsilon(p_0){(-1)^{N}\over
N!}\delta^{(N)}(x-\lambda)\biggr\rbrace x^{2n-i+l+m\varepsilon}\ \ln^{r'} x\qquad(3.46)}$$As in (3.22), we
introduce the representation
$$\ln^{r'} x=\lim_{{\widehat{\varepsilon}}=0}\ {(-1)^{r'}\over
{\widehat{\varepsilon}}^{r'}}\sum_{s'=0}^{r'}C_{r'}^{s'}(-1)^{s'}(x)^{s'{\widehat{\varepsilon}}}\eqno(3.47)$$and
interchange the sum on $s'$, which is finite, with the integration on $x$. In the limit
$\lambda=0$, we get first the expression$^{7}$, $${(-1)^{r'}\over
{\widehat{\varepsilon}}^{r'}}\sum_{s'=0}^{r'}C_{r'}^{s'}(-1)^{s'}\ {-\sin(\pi m\varepsilon)\over
N-(2n-i+l)+m\varepsilon+s'{\widehat{\varepsilon}}}\ \ +\ {\cal{O}}(\lambda)\eqno(3.48)$$At
$N\geq
2n-i+l+1$, such a factor admits an $s'{\widehat{\varepsilon}}$ -Taylor series expansion, so that relying on
arithmetical identities
(3.28) and (3.29), the ${\widehat{\varepsilon}}=0$-limit of (3.48) is readily obtained to be given by
$$\left({\sin(\pi m\varepsilon)\over
N-(2n-i+l)+m\varepsilon+s'{\widehat{\varepsilon}}}\right)^{(r')}_{|_{s'{\widehat{\varepsilon}}\ =\
0}}=(-1)^{r'}r'!\ {\sin(\pi m\varepsilon)\over \left(N-(2n-i+l)+m\varepsilon\right)^{r'+1}} \eqno(3.49)$$ At its
turn, (3.49) admits itself a Taylor series expansion in the variable $m\varepsilon$. 
Since the whole expression (3.46) factors out a global factor
of
$${1\over
\varepsilon^{2(n-i)+k}}\sum_{m=0}^{2(n-i)+k}\ C_{2(n-i)+k}^m\ (-1)^m\
F_{km}^{(l)}(m\varepsilon,0)\eqno(3.50)$$the $\varepsilon=0$-limit of
(3.46) is finite in view, again, of
arithmetical identities
(3.28) and (3.29), and extra factors of type (3.45) are {\it{msf}}-preserving too.\bigskip
We thus reach the conclusion that generic type (3.11) integrals are {\it{msf}}. Now, getting back to the
mass singularity issue of $(N,N';0)$ self energy diagrams, it is immediate to realize that all of the trace
factors
(3.8)-(3.10) only involve $x$($x'$),${\sqrt{1-x}}$(${\sqrt{1-x'}}$), and $[\epsilon(p_0)\ln
X]$($[\epsilon(p'_0)\ln
X' ]$) {\it{msf}}-preserving dependences (some of them will be treated in full details in next Section 5), as, for
example, the most involved piece of (3.10) $$\displaylines{4P^2P'^2\Sigma_R(P)\!\cdot\!\Sigma_R(P')=4p(p_0+q){xx'\over 1-x'}(m^2)^2Q_0({p_0\over p})\ 
 \  {\sqrt{1-x'}}\ Q_0({p'_0\over p'})\cr\hfill -4p{xx'\over 1-x'}(m^2)^2Q_1({p_0\over p})\ Q_1({p'_0\over
p'})\biggl\lbrace p(1-x')+{q\over 2}({p\over q}+{q\over
 p}){x'}+\left(q{\sqrt{1-x}}-{p\over 2}x\right)\biggr\rbrace\qquad}$$\smallskip Since the
same
 analysis can be carried through in the case of negative energies, we can
 conclude that $(N,N';0)$ self energy diagrams have collinear singularity free imaginary parts.
 \bigskip\bigskip\bigskip
 
 {\bf{IV. Diagrams of type $(N,N';1)$, with one effective vertex}}\bigskip

 Using (3.4), the trace factors associated with diagrams of type $(N,N';1)$, depicted in Fig.2, are easily
obtained
 to  be, $$8\ ({\widehat {K}}\!\cdot\! P)({\widehat {K}}\!\cdot\! P')\eqno(4.1)$$
$$16m^2\ ({\widehat {K}}\!\cdot\!\! P)({\widehat {K}}\!\cdot\!\! P')-8P^2{\widehat {K}}\!\cdot\!\!P'{\widehat
{K}}\!\cdot\!\Sigma_R(P) \eqno(4.2)$$
$$(4.2)\left(P\leftrightarrow P'\right)\eqno(4.2')$$
$$\displaylines{8(2m^2)^2\ ({\widehat {K}}\!\cdot\! P)({\widehat {K}}\!\cdot\! P')
-8(2m^2)\left(P^2{\widehat
{K}}\!\cdot\!\!P'{\widehat {K}}\!\cdot\!\Sigma_R(P)+(P\leftrightarrow P')\right)\cr\hfill +8P^2P'^2{\widehat
{K}}\!\cdot\!\Sigma_R(P){\widehat {K}}\!\cdot\!\Sigma_R(P')\qquad (4.3)}$$In $(N,N';1)$-type diagrams, each
of the trace factors (4.1)-(4.3) must be integrated over ${\widehat {K}}$ with the "measure",
$$m^2\!\int{{\rm d}{\widehat
K}\over 4\pi}\ {1\over ({\widehat K}\!\cdot\! P+i\epsilon)({\widehat
K}\!\cdot\! P'+i\epsilon)}\eqno(4.4)$$Then the above trace factors yield respectively
$$8m^2\eqno(4.5)$$
$$ 16m^4-8P^2\Sigma^2_R(P)\eqno(4.6)$$
$$(4.6)\left(P\leftrightarrow P'\right)\eqno(4.6')$$
$$8m^2\left((2m^2)^2-2\left(P^2\Sigma^2_R(P)+(P\rightarrow P')\right)+P^2{P'}^2\int{{\rm d}{\widehat
K}\over 4\pi}\ {{\widehat {K}}\!\cdot\!\Sigma_R(P)\over {\widehat K}\!\cdot\! P+i\epsilon}
{{\widehat {K}}\!\cdot\!\Sigma_R(P')\over {\widehat K}\!\cdot\! P'+i\epsilon}\right)\eqno(4.7)$$One may observe
that the trace factors of $(N,N';0)$ diagrams are more involved than those attached to
diagrams of type $(N,N';1)$. In particular, it should be clear that at the exception of the last term of (4.7),
all
of the factors appearing in (4.5)-(4.7) will preserve the {\it{msf}}
character of type (3.11) integrals.\medskip That is, the whole mass singularity issue of $(N,N';1)$ contributions
is entirely in the
incidence, upon generic
type (3.11) integrals, of the very function
$$\int{{\rm d}{\widehat
K}\over 4\pi}\ \ {{\widehat {K}}\!\cdot\!\Sigma_R(P)\over {\widehat K}\!\cdot\! P+i\epsilon}\ 
{{\widehat {K}}\!\cdot\!\Sigma_R(P')\over {\widehat K}\!\cdot\! P'+i\epsilon}\eqno(4.8)$$As it stands, (4.8)
can
be
calculated with the help of the three {\it{angular identities}},
$$\int{{\rm d}{\widehat
K}\over 4\pi}\ {{\widehat K}^0{\widehat K}^0\over ({\widehat K}\!\cdot\! R+i\epsilon)^2}={1\over R^2+i\epsilon
r_0}\eqno(4.9)$$
$$\int{{\rm d}{\widehat
K}\over 4\pi}\ {{\widehat K}^0{\widehat K}^i\over ({\widehat K}\!\cdot\! R+i\epsilon)^2}={\widehat
r}^i\left({-1\over
2r^2}\ln({r_0+r\over r_0-r})+{r_0\over r}{1\over R^2+i\epsilon r_0}\right)\eqno(4.10)$$
$$\int{{\rm d}{\widehat
K}\over 4\pi}\ {{\widehat K}^i{\widehat K}^j\over ({\widehat K}\!\cdot\! R+i\epsilon)^2}=-{g^{ij}\over
r^2}Q_1({r_0\over r})-{\widehat r}^i{\widehat r}^j\left({3\over
r^2}Q_1({r_0\over r})-{1\over R^2+i\epsilon
r_0}\right)
\eqno(4.11)$$where a Feynman
parameter, $s$, has been introduced so as to re-write (4.8) as
$$\Sigma_\mu(P)\Sigma(P')_\nu\ \int_0^1{\rm{d}}s\int{{\rm d}{\widehat
K}\over 4\pi}\ {{\widehat K}^\mu{\widehat K}^\nu\over ({\widehat K}\!\cdot \! R(s)+i\epsilon)^2}
\eqno(4.12)$$with,
$$R(s)=P+sQ\eqno(4.13)$$At this point, and though not immediately relevant to our concern, the following remark
may
be in order.\medskip Some
years ago, the use of a Feynman parametrization in Thermal Quantum Field Theories has been
questioned$^{20}$. Feynman
parametrization was suspected delicate, using, for example, bare propagator determinations different from
the usual $+i\epsilon$-Feynman's one. Of course, passing from (4.8) to (4.12), this situation is not
encountered, but the difficulty may come about, in particular in a real time formalism using
$Retarded/Advanced$ propagator prescriptions. The solution to this difficulty has been given in Ref.21. Later
on, it has even been suggested that using a Feynman parametrization in a hot quantum field context could lead to
non gauge invariant results$^{22}$. This latter statement however was erroneous, based on incorrect
calculations,
and
indeed, taking the modification of Ref.21 into account, it must be stated that there is definitely no problem in
dealing with
Feynman parametrizations in non zero Temperature Quantum Field Theories.\medskip Getting back to (4.8), a
shortcut
to its calculation consists in writing, 
$${{\widehat K}\!\cdot\!\Sigma_R(P)\over {\widehat K}\!\cdot\!P+i\epsilon}={m^2\over p^2}Q_1({p_0\over p})
+{m^2\over p}\left({p_0\over p}-{1\over 2}{P^2\over p^2}\ln X\right)\ {1\over {\widehat
K}\!\cdot\!P+i\epsilon}\eqno(4.14)$$from which a remarkable relation may be deduced,
$$m^2\!\int{{\rm d}{\widehat
K}\over 4\pi}\ \ {{\widehat {K}}\!\cdot\!\Sigma_R(P)\over {\widehat K}\!\cdot\!
P+i\epsilon}=\Sigma^2_R(P)\eqno(4.15)$$and likewise, in obvious notations,
$$\displaylines{(4.8)={m^2\over p^2}Q_1{m^2\over p'^2}Q'_1+\left({m^2\over p^2}Q_1{m^2\over p'}({p'_0\over
p'}-{1\over
 2}{P'^2\over p'^2}\ln X' ){1\over 2p'}\ln X'++(P\leftrightarrow P')\right)\cr\hfill 
 +{m^2\over p}({p_0\over p}-{1\over 2}{P^2\over p^2}\ln X )\ {m^2\over p'}({p'_0\over p'}-{1\over
 2}{P'^2\over p'^2}\ln X' )\ {1\over 2Q\!\cdot\!P+i\epsilon q_0}\ln{P'^2+i\epsilon p'_0\over P^2+i\epsilon p_0} 
 \qquad(4.16)}$$where identity (4.9) only, has been used. Noting that $2Q\!\cdot\!P=P'^2-P^2$, one recovers in
 (4.16) the full original symmetry of (4.8) under the exchange $P\leftrightarrow P'$. Since the terms
appearing in (4.16) just redefine the integer numbers $k(k')$, $n(n')$ and $2(n-i)+k\ (2(n'-i')+k')$, they leave
totally unaffected the {\it{msf}} structures of previous Section 3. The only new
feature is the factor $(2Q\!\cdot\! P)^{-1}\ln{P'^2/ P^2}$. As observed in
Ref.13 for the topology $(1,1;1)$, this factor is reminiscent of the
collinear singularity
plaguing $(N,N';1)$ diagrams at the light cone, when an $RP$ treatment of the problem is adopted. \smallskip In
the
end, recalling that (4.8) comes out affected with a multiplicative factor of $8m^2P^2P'^2$, this means that
expressions $$\displaylines{\int {{\rm d}^4P\over (2\pi)^4}(1-2n_F(p_0))\ \left({(-{P'}^2\Sigma^2_R(P'))^{n'}\over ({P'}^2+i\epsilon
p'_0)^{N'}}\right)\cr\hfill {\rm{disc}}\left({(-{P}^2\Sigma^2_R(P))^{n}\over ({P}^2+i\epsilon
p_0)^{N}}\right)\ \ \int{{\rm d}{\widehat
K}\over 4\pi}\ {{\widehat {K}}\!\cdot\!\Sigma_R(P)\over {\widehat
K}\!\cdot\! P+i\epsilon}\ {{\widehat {K}}\!\cdot\!\Sigma_R(P')\over {\widehat K}\!\cdot\!
P'+i\epsilon}\qquad(4.17)}$$do have {\it{msf}} imaginary parts, if integrals 
$$\int
{{\rm d}^4P\over (2\pi)^4}(1-2n_F(p_0))\ {(-{P'}^2\Sigma^2_R(P'))^{n'}\over ({P'}^2+i\epsilon
p'_0)^{N'}}\  {\rm{disc}}{(-{P}^2\Sigma^2_R(P))^{n}\over ({P}^2+i\epsilon
p_0)^{N}}\ {1\over 2Q\!\cdot\!P+i\epsilon q_0}\ln{P'^2+i\epsilon p'_0\over P^2+i\epsilon p_0}\eqno(4.18)$$have
{\it{msf}} imaginary parts either. That it is
so can be demonstrated quite easily. However, a byproduct of the next section will provide this statement with a
systematic
derivation,
so that
we can here content ourselves with a heuristic, still instructive argument.\smallskip The potential collinear
singularity
due to the $HTL$ vertex comes from the factor $(1/2Q\!\cdot\!P)$, as $Q\!\cdot\!P$ reaches zero. For example,
in
the $RP$ calculation of Sec.2, we learn out of Eqs.(2.12)-(2.14), that the collinear singularity expression
effectively
involves a $\delta(P\!\cdot\!{Q})$ constraint. Now, as $Q\!\cdot\!P$ tends to zero, one has indeed
$${1\over 2Q\!\cdot\!P+i\epsilon q_0}\ln{P^2+2Q\!\cdot\!P+i\epsilon p'_0\over P^2+i\epsilon p_0}\simeq {1\over
P^2+i\epsilon p_0}\eqno(4.19)$$ and this light cone potentially singular behaviour obviously gets mixed with
partial effective propagator $S^{(N)}_R(P)$ own light cone
potentially singular
behaviour,
$$P^2S^{(N)}_R(P)=P^2{i\rlap / \!P\left(\rlap /\!\Sigma_R(P)\rlap /\!
P\right)^{N}\over (P^2+i\epsilon p_0)^{N+1}}\eqno(4.20)$$the
whole just boiling down to a simple shift of power,
$$P^2{1\over (P^2+i\epsilon p_0)^{N+1}}\  \longmapsto\ P^2{1\over (P^2+i\epsilon p_0)^{N+2}}$$From previous Sec.3, Eqs.
(3.43)-(3.44), we know that the overall detailed balance compensation of mass
singularities is preserved by such a shift, and this is how we can see that $(N,N';1)$ contributions to the
soft
real
photon emission rate
are {\it{msf}}.\smallskip This generalizes to any $(N,N';1)$ contribution,
the observation first made in Ref.13, Sec.5, for the diagram $(1,1;1)$, and simply enforces the conclusion we
drew
then, that $HTL$ vertex collinear singularities should not be desentangled from
partial
effective propagator mass singularities, as they
all
mix up into structural patterns which grant their overall compensations. In an $RP$ resummation scheme,
unfortunately, a dissociation of Eqs.(4.19) and (4.20) potentially singular behaviours is achieved right from the
onset. There, in effect, the sum over $N$ being performed before the integration on $p_0$, partial effective
propagators $S^{(N)}_R(P)$, get replaced by full effective ones, ${}^\star S_R(P)$, whose poles, contrarily to
$S^{(N)}_R(P)$-poles, are no longer lightlike
at $P^2\simeq 0$,
but timelike, at $p_0=\pm \omega_s(p)$$^1$. It results that the light cone singular behaviour of (4.19) remains
isolated, with no other singular behaviour to cancel with.

\bigskip\bigskip\bigskip
{\bf{V. Two effective vertex diagrams $(N,N';2)$}}
\bigskip
We now turn to the analysis of $(N,N';2)$ topologies depicted in Fig.3, which are the most important to consider,
the famous collinear problem of hot $QCD$ being induced by these double effective vertex insertions.\smallskip In
Ref.13, it was shown that $(1,0;2)$ is singularity free. While an encouraging result, it would certainly be
preposterous to take it for granted that the property trivially extends to any $(N,N';2)$ diagram, and in our
opinion, this is why the present analysis had to be undertaken.\smallskip
The contribution to $\Pi_R(Q)$ of a diagram $(N,N';2)$ reads,
$$\displaylines{ \Pi^{(N,N';2)}_R(Q)= ie^2m^4\int {{\rm d}^4P\over
(2\pi)^4}(1-2n_F(p_0))\int{{\rm d}{\widehat
K}\over 4\pi}\int{{\rm d}{\widehat
K'}\over 4\pi}\ {\widehat
K}\!\cdot\!{\widehat
K'}  \cr\hfill
{\rm
disc}\ \ Tr\left(\rlap / \!P{\left(\rlap /\!\Sigma_R(P)\rlap /\!
P\right)^N\over (P^2+i\epsilon p_0)^{N+1}}\ {{\rlap / \!\widehat K}}\ \rlap / \!P'{\left(\rlap
/\!\Sigma_R(P')\rlap
/\! P'\right)^{N'}\over (P'^2+i\epsilon p'_0)^{N'+1}}\ {{\rlap / \!\widehat
K'}}\right)
\cr\hfill\times {1\over ({\widehat K}\!\cdot\! P+i\epsilon)({\widehat
K}\!\cdot\! P'+i\epsilon)({\widehat K'}\!\cdot\! P+i\epsilon)({\widehat
K'}\!\cdot\! P'+i\epsilon)}\qquad(5.1)} $$Expanding the $\left(\rlap /\!\Sigma_R(P)\rlap /\!
P\right)^N$ factors as in (3.4), four types of traces come about,
$$4\left(2{\widehat K}\!\cdot\! P{\widehat K'}\!\cdot\! P'-{\widehat K}\!\cdot\! {\widehat
K'}P\!\cdot\!P'\right)\eqno(5.2)$$
$$8m^2\left(2{\widehat K}\!\cdot\! P{\widehat K'}\!\cdot\! P'-{\widehat K}\!\cdot\! {\widehat
K'}P\!\cdot\!P'\right)-4P^2\left(2{\widehat K}\!\cdot\! \Sigma{\widehat K'}\!\cdot\! P'-{\widehat K}\!\cdot\!
{\widehat K'}P'\!\cdot\!\Sigma\right)\eqno(5.3)$$
$$8m^2\left(2{\widehat K}\!\cdot\! P{\widehat K'}\!\cdot\! P'-{\widehat K}\!\cdot\! {\widehat
K'}P\!\cdot\!P'\right)-4P'^2\left(2{\widehat K}\!\cdot\! \Sigma'{\widehat K'}\!\cdot\! P-{\widehat K}\!\cdot\!
{\widehat K'}P\!\cdot\!\Sigma'\right)\eqno(5.4)$$
$$\displaylines{(2m^2)^2\left(8{\widehat K}\!\cdot\! P{\widehat K'}\!\cdot\! P'-4{\widehat K}\!\cdot\!
{\widehat K'}P\!\cdot\!P'\right)-8m^2P'^2\left(2{\widehat K}\!\cdot\! \Sigma'{\widehat K'}\!\cdot\! P-{\widehat
K}\!\cdot\!
{\widehat K'}P\!\cdot\!\Sigma'\right)\cr\hfill
-8m^2P^2\left(2{\widehat K}\!\cdot\! \Sigma{\widehat K'}\!\cdot\! P'-{\widehat K}\!\cdot\!
{\widehat K'}P'\!\cdot\!\Sigma\right)+4P^2P'^2\left(2{\widehat K}\!\cdot\! \Sigma{\widehat K'}\!\cdot\!
\Sigma'-{\widehat K}\!\cdot\!
{\widehat K'}\Sigma\!\cdot\!\Sigma'\right)
\qquad(5.5)}$$Integrated on both light-like vectors ${\widehat K},{\widehat K'}$, the first trace (5.2) yields
the
expression
$${8\over m^4}\Sigma\!\cdot\!\Sigma'-4P\!\cdot\!P'W_2(P,P')\eqno(5.6)$$where
$W_2(P,P')$ is the double vertex function met in Ref.13,
$$W_2(P,P')=\int{{\rm d}{\widehat K}\over 4\pi}\int{{\rm d}{\widehat
K'}\over 4\pi}\ {({\widehat
K}\!\cdot\!{\widehat
K'})^2\over ({\widehat K}\!\cdot\! P+i\epsilon)({\widehat
K}\!\cdot\! P'+i\epsilon)({\widehat K'}\!\cdot\! P+i\epsilon)({\widehat
K'}\!\cdot\! P'+i\epsilon)}\eqno(5.7)$$The second trace (5.3) gives
$$2m^2\times (5.6)-4P^2\biggl\lbrace{2\over m^2}\int{{\rm d}{\widehat
K}\over 4\pi}\ {({\widehat
K}\!\cdot\!\Sigma)^2\over ({\widehat K}\!\cdot\! P+i\epsilon)({\widehat
K}\!\cdot\! P'+i\epsilon)}
 -P'\!\cdot\!\Sigma\ W_2(P,P')\biggr\rbrace\eqno(5.8)$$The third trace (5.4) gives the same as (5.8) with $P$ and
$P'$ interchanged,
$$2m^2\times (5.6)-4P'^2\biggl\lbrace{2\over m^2}\int{{\rm d}{\widehat
K}\over 4\pi}\ {({\widehat
K}\!\cdot\!\Sigma')^2\over ({\widehat K}\!\cdot\! P+i\epsilon)({\widehat
K}\!\cdot\! P'+i\epsilon)}
 -P\!\cdot\!\Sigma'\ W_2(P,P')\biggr\rbrace\eqno(5.9)$$Eventually, the fourth trace (5.5) yields
$$\displaylines{-(2m^2)^2\times (5.6)+2m^2\times (5.8)+2m^2\times (5.9) -4P^2P'^2\ \Sigma\!\cdot\!\Sigma'\
W_2(P,P')\cr\hfill +8P^2P'^2\int{{\rm d}{\widehat K}\over 4\pi}\int{{\rm d}{\widehat
K'}\over 4\pi} {({\widehat
K}\!\cdot\!{\widehat
K'})({\widehat
K}\!\cdot\!\Sigma)({\widehat
K'}\!\cdot\!\Sigma')\over ({\widehat K}\!\cdot\! P+i\epsilon)({\widehat
K}\!\cdot\! P'+i\epsilon)({\widehat K'}\!\cdot\! P+i\epsilon)({\widehat
K'}\!\cdot\! P'+i\epsilon)} \qquad(5.10)}$$To summarize, the
whole expression (5.1) reads $$\displaylines{ \Pi^{(N,N';2)}_R(Q)= ie^2m^4\int {{\rm d}^4P\over
(2\pi)^4}(1-2n_F(p_0))\ {1\over (P'^2+i\epsilon p'_0)^{N'+1}} \cr\hfill
{\rm
disc}\ {1\over (P^2+i\epsilon p_0)^{N+1}}\ \bigl\lbrace b_Nb_{N'}\times (5.6)+a_Nb_{N'}\times
(5.8)+b_Na_{N'}\times
(5.9)+a_Na_{N'}\times (5.10)\bigr\rbrace
\qquad(5.11)} $$where the coefficients $a_N, b_N$, polynomials of degree
$j_M(N)$ and $j_M(N)+1$ in the variable $-P^2\Sigma^2/m^4$, respectively, are given in (3.5) and (3.6).
That is, one would have again to investigate the incidence upon generic type (3.11) integrals, of the new
multiplicative functions appearing through
(5.6)-(5.10). For example, we quote that an expression like (5.6) will contribute
$(N,N';2)$ a quantity,
$$\displaylines{
(m^2)^{N+N'}\ \sum_{j=0}^{j_M}C_{N-1}^{2j+1}\ \sum_{j'=0}^{j'_M}C_{N'-1}^{2j'+1}\ \sum_{n=0}^{j}C_{j}^{n-1}\ 
\sum_{n' =0}^{j'}C_{j'}^{n'-1}\ \times \ ie^2m^4\int {{\rm d}^4P\over
(2\pi)^4}(1-2n_F(p_0))\cr {1\over (m^4)^{n+n'}}\ {(-P'^2\Sigma'^2)^{n'}\over (P'^2+i\epsilon p'_0)^{N'+1}}\ 
{\rm
disc}\ {(-P^2\Sigma^2)^{n}\over (P^2+i\epsilon p_0)^{N+1}} \ \left( {8\over
m^4}\Sigma\!\cdot\!\Sigma'-4P\!\cdot\!P'W_2(P,P')\right)\qquad(5.12)}$$where the sums over $j$ and $n$, which are
finite, have
been
interchanged with the integral on $P$.
\medskip Actually, things may be further reduced, and this helps identifying the new mutiplicative functions that
come out to be specific to the double effective vertex diagrams. To do so, we can make use of the relation
$$\displaylines{\int{{\rm d}{\widehat K}\over 4\pi}{({\widehat K}\!\!\cdot\!\!\Sigma)^2\over ({\widehat
K}\!\!\cdot\!\!
P+i\epsilon)({\widehat
K}\!\cdot\! P'+i\epsilon)}={1\over p^2}Q_1({p_0\over p})\Sigma\!\cdot\!\Sigma'\cr\hfill
+{m^2\over p}\left({p_0\over p}-{1\over 2}{P^2\over p^2}\ln X\right){m^2\over p^2}Q_1({p_0\over p}){1\over
2p'}\ln
X'\cr\hfill
+\left({m^2\over p}\left({p_0\over p}-{1\over 2}{P^2\over p^2}\ln X\right)\right)^2\ {1\over
2Q\!\cdot\!P+i\epsilon q_0}\ln{P^2+2Q\!\cdot\!P+i\epsilon p'_0\over P^2+i\epsilon p_0}\qquad(5.13)}$$ and of a
similar one, with $P'$ and $P$ interchanged, and likewise,
 $$\displaylines{\int{{\rm d}{\widehat
K}\over 4\pi}\int{{\rm d}{\widehat
K'}\over 4\pi} {({\widehat
K}\!\cdot\!{\widehat
K'})({\widehat
K}\!\cdot\!\Sigma)({\widehat
K'}\!\cdot\!\Sigma')\over ({\widehat K}\!\cdot\! P+i\epsilon)({\widehat
K}\!\cdot\! P'+i\epsilon)({\widehat K'}\!\cdot\! P+i\epsilon)({\widehat
K'}\!\cdot\! P'+i\epsilon)}={1\over p^2}Q_1({p_0\over p}){1\over p'^2}Q_1({p'_0\over
p'})\Sigma\!\cdot\!\Sigma'\cr\hfill
+{m^2\over p^2}Q_1({p_0\over p}){m^2\over p'^2}Q_1({p'_0\over p'})\left({1\over p}({p_0\over p}-{1\over
2}{P^2\over p^2}\ln X){1\over 2p'}\ln X'\ +\ (P\leftrightarrow P')\right)\cr\hfill
+\left({1\over p'^2}Q_1({p'_0\over p'})\left({m^2\over p}({p_0\over p}-{1\over 2}{P^2\over
p^2}\ln X)\right)^2\ {1\over 2Q\!\cdot\!P+i\epsilon q_0}\ln{P^2+2Q\!\cdot\!P+i\epsilon p'_0\over P^2+i\epsilon
p_0}+\ (P\leftrightarrow P')\right)\cr\hfill 
+\left({m^2\over p}({p_0\over p}-{1\over 2}{P^2\over
p^2}\ln X)\right)\left({m^2\over p'}({p'_0\over p'}-{1\over 2}{P'^2\over
p'^2}\ln X')\right)\ W_1(P,P')\qquad(5.14)}$$where $W_1(P,P')$ is another double effective vertex
function, not encountered in Ref.13,
$$W_1(P,P')=\int{{\rm d}{\widehat K}\over 4\pi}\int{{\rm d}{\widehat
K'}\over 4\pi}\ {({\widehat
K}\!\cdot\!{\widehat
K'})^1\over ({\widehat K}\!\cdot\! P+i\epsilon)({\widehat
K}\!\cdot\! P'+i\epsilon)({\widehat K'}\!\cdot\! P+i\epsilon)({\widehat
K'}\!\cdot\! P'+i\epsilon)}\eqno(5.15)$$Deriving (5.13) and (5.14), identities (4.9) and (4.14) only, have been
used. As expected
on the basis of general gauge invariance arguments$^{15}$, (5.13) and (5.14) entail some potential collinear
structures similar those found in the case of one effective vertex diagrams, (4.16). Now, a
comparison with the previous cases of $(N,N';0)$ and $(N,N';1)$, also allows to identify with $W_1$ and $W_2$,
the new extra mutiplicative functions that come out to be specific to the double effective vertex diagrams,
$(N,N';2)$.\medskip
 But it seems difficult to proceed further : As shown in the Appendix, an explicit calculation of functions
$W_1$ and $W_2$
is doable, relying this time, on the full set of angular identities (4.9)-(4.11). Results, however, come out so
cumbersome that controlling the ensuing integrals on $x'$ and on $x$, is rendered extremely hazardous. In order
to proof
that
(5.11) does
have an {\it{msf}} imaginary part, we must therefore proceed differently, and construct a proof by
induction.\bigskip
We consider the contribution to $\Pi_R(Q)$ of the diagram $(N+2,N';2)$. It is, 
$$\displaylines{\Pi_R^{(N+2,N';2)}(Q)= 8ie^2m^4\int {p^2{\rm d}p\over
(2\pi)^2}\sum_{\epsilon(p_0)}\int_0^1 {{\rm d}x\over
2\pi}\   {p_0(x)\over 2T}\  {\rm{disc}} {1\over (P^2)_R^{N+3}}\cr\hfill\times\int_0^{x'_M} {{\rm
d}x'\over (P'^2)_R^{N'+1}}\int_{{\widehat K},{\widehat
K'}}\ \ {\widehat
K}\!\cdot\!{\widehat
K'}\    { Tr\left(\rlap / \!P\left(\rlap /\!\Sigma_R\rlap /\!
P\right)^{N+2}\ {{\rlap / \!\widehat K}}\ \rlap / \!P'\left(\rlap
/\!\Sigma'_R\rlap
/\! P'\right)^{N'} {{\rlap / \!\widehat
K'}}\right)\over ({\widehat K}\!\cdot\! P)_R({\widehat
K}\!\cdot\! P')_R({\widehat K'}\!\cdot\! P)_R({\widehat
K'}\!\cdot\! P')_R}\qquad(5.16)}$$where some obvious shorthand notations have been introduced so as to
alleviate
too large
expressions,$$\displaylines{\int{{\rm d}{\widehat K}\over 4\pi}\int{{\rm d}{\widehat
K'}\over 4\pi}\equiv \int_{{\widehat K},{\widehat
K'}}\ ,\ \ \ {1\over P^2+i\epsilon p_0}\equiv {1\over (P^2)_R}\ ,\ \ \ \Sigma_R(P)\equiv\Sigma_R\ ,\ \ \
\Sigma_R(P)'\equiv\Sigma'_R\cr\hfill {1\over P'^2+i\epsilon p'_0}\equiv {1\over (P'^2)_R}\ ,\ \ \
{1\over{\widehat
K} \!\cdot\!P+i\epsilon p_0}\equiv {1\over ({\widehat
K} \!\cdot\!P)_R}\ ,\ .. 
\qquad(5.17)}$$
Having,
$$\left(\rlap /\!\Sigma_R\rlap /\!
P\right)^{2}=2m^2\ \rlap /\!\Sigma_R\rlap /\!
P-P^2\Sigma^2_RI\!\!I_4\eqno(5.18)$$Eq.(5.16) may be written as,
$$\displaylines{8ie^2m^4\int {p^2{\rm d}p\over
(2\pi)^2}\sum_{\epsilon(p_0)}\int_0^1 {{\rm d}x\over
2\pi}\   {p_0(x)\over 2T}\biggl\lbrace {\rm{disc}}\  {2m^2\over (P^2)_R^{N+3}}\cr\hfill\times\int_0^{x'_M} {{\rm
d}x'\over (P'^2)_R^{N'+1}}\int_{{\widehat K},{\widehat
K'}}\ \ {\widehat
K}\!\cdot\!{\widehat
K'}\  { Tr\left(..\left(\rlap /\!\Sigma_R\rlap /\!
P\right)^{N+1}\ ..\right)\over ({\widehat K}\!\cdot\! P)_R\ . .\ ({\widehat
K'}\!\cdot\! P')_R}\cr\hfill
-{\rm{disc}}\  {\Sigma_R^2\over (P^2)_R^{N+2}}\int_0^{x'_M}\!\! {{\rm
d}x'\over (P'^2)_R^{N'+1}}\int_{{\widehat K},{\widehat
K'}}\ \ {\widehat
K}\!\cdot\!{\widehat
K'}\   { Tr\left(..\left(\rlap /\!\Sigma_R\rlap /\!
P\right)^{N}\ ...\right)\over ({\widehat K}\!\cdot\! P)_R\ . .\ ({\widehat
K'}\!\cdot\! P')_R}\biggr\rbrace\qquad(5.19)}$$where the dots stand for all of those factors which are left the
same as in (5.16).
Next, we can form the difference of (5.16) with the first term of (5.19), obtaining
$$\displaylines{8ie^2m^4\int {p^2{\rm d}p\over
(2\pi)^2}\sum_{\epsilon(p_0)}\int_0^1 {{\rm d}x\over
2\pi}\   {p_0(x)\over 2T}\ {\rm{disc}}\  {1\over (P^2)_R^{N+3}}\cr\hfill\times\int_0^{x'_M}\!\! {{\rm
d}x'\over (P'^2)_R^{N'+1}}\int_{{\widehat K},{\widehat
K'}}\ \ {\widehat
K}\!\cdot\!{\widehat
K'}\   { Tr\left(..\Bigl\lbrace\left(\rlap /\!\Sigma_R\rlap /\!
P\right)^{N+2}-2m^2\left(\rlap /\!\Sigma_R\rlap /\!
P\right)^{N+1}\Bigr\rbrace\ ..\right)\over ({\widehat K}\!\cdot\! P)_R\ . .\ ({\widehat
K'}\!\cdot\! P')_R}\qquad(5.20)}$$This difference is, of course, also given by,
$$\displaylines{-8ie^2m^4\int {p^2{\rm d}p\over
(2\pi)^2}\sum_{\epsilon(p_0)}\int_0^1 {{\rm d}x\over
2\pi}\   {p_0(x)\over 2T}\ {\rm{disc}}\  {\Sigma^2\over (P^2)_R^{N+2}}\cr\hfill\times\int_0^{x'_M}\!\! {{\rm
d}x'\over (P'^2)_R^{N'+1}}\int_{{\widehat K},{\widehat
K'}}\ \ {\widehat
K}\!\cdot\!{\widehat
K'}\   { Tr\left(..\left(\rlap /\!\Sigma_R\rlap /\!
P\right)^{N}\ ..\right)\over ({\widehat K}\!\cdot\! P)_R\ . .\ ({\widehat
K'}\!\cdot\! P')_R}\qquad(5.21)}$$Our induction hypothesis is that ${{(N+1,N';2)}}$ is endowed with an
{\it{msf}} imaginary part, that is, the expression
$$\displaylines{8ie^2m^4\int {p^2{\rm d}p\over
(2\pi)^2}\sum_{\epsilon(p_0)}\int_0^1 {{\rm d}x\over
2\pi}\   {p_0(x)\over 2T}\ {\rm{disc}}\  {1\over (P^2)_R^{N+2}}\cr\hfill\times\int_0^{x'_M}\!\! {{\rm
d}x'\over (P'^2)_R^{N'+1}}\int_{{\widehat K},{\widehat K'}}\ \ {\widehat
K}\!\cdot\!{\widehat
K'} \  { Tr\left(..\left(\rlap /\!\Sigma_R\rlap /\!
P\right)^{N+1}\ ..\right)\over ({\widehat K}\!\cdot\! P)_R\ . .\ ({\widehat
K'}\!\cdot\! P')_R}\qquad(5.22)}$$\medskip We will need the following result: Let there be $S^{(2)}(q,p;p_0)$
some
function,
such that, for all integers $k$ and $l$, and all integers $i$ and $n$, with $ 0\leq n-i$, the
following finite sum of integrals, $$\displaylines{{1\over
\varepsilon^{2(n-i)+k}}\sum_{m=0}^{2(n-i)+k}C^m_{2(n-i)+k}{(-1)^m\
{F_{km}^{(l)}(m\varepsilon,0)} } \int_0^1{\rm{d}}x\ \Bigl\lbrace  {\bf{P}}{{\cal{I}}m\left(e^{i\pi
m\varepsilon}S^{(2)}(q,p;p_0)\right)\over
(x-\lambda)^{N+1}}\cr\hfill +
\pi\epsilon(p_0){(-1)^{N}\over N!}\delta^{(N)}(x-\lambda)\ {\cal{R}}e \left(e^{i\pi
m\varepsilon}S^{(2)}(q,p;p_0)\right)\Bigr\rbrace x^{2n-i+l+m\varepsilon}\qquad(5.23)}$$has
{\it{msf}} imaginary part in the limits $\lambda=0$ and $\varepsilon=0$, with no further specifications
required$^7$. Then we claim that so is the case of the finite sum
of integrals, $$\displaylines{{1\over
\varepsilon^{2(n-i)+k}}\sum_{m=0}^{2(n-i)+k}C^m_{2(n-i)+k}{(-1)^m {F_{km}^{(l)}(m\varepsilon,0)} }
\int_0^1{\rm{d}}x \Bigl\lbrace  {\bf{P}}{{\cal{I}}m\left(e^{i\pi
m\varepsilon}\  S^{(2)}(q,p;p_0)\times \Sigma_R^2(P)\right)\over
(x-\lambda)^{N+1}}\cr\hfill +
\pi\epsilon(p_0){(-1)^{N}\over
N!}\delta^{(N)}(x-\lambda)\ {\cal{R}}e \left(e^{i\pi m\varepsilon}\
S^{(2)}(q,p;p_0)\times\Sigma_R^2(P)\right)\Bigr\rbrace x^{2n-i+l+m\varepsilon}
\qquad(5.24)}$$with, $$\Sigma_R^2(P)={m^4\over
p^2}\left(-{1}+{x\over 4}[\epsilon(p_0)\ln X]^2+{\sqrt{1-x}}[\epsilon(p_0)\ln
X]\right)\eqno(5.25)$$Before we proceed further, the relation of structural patterns
(5.23) (and (5.24)),
with a general term $(N,N';2)$ is worth making explicit. This is achieved by noting that the imaginary part
of
$(N,N';2)$ can be written
as, $$8e^2m^4\int {p^2{\rm d}p\over
(2\pi)^2}\sum_{\epsilon(p_0)=\pm 1}\int_0^1 {{\rm d}x\over
2\pi}\   {p_0(x)\over 2T}\ {\rm{disc}}_{P}\  {
a_{N}S^{(2)}_a+b_{N}S^{(2)}_b\over
(P^2)_R^{N+1}}\eqno(5.26)$$with $p_0(x)=\epsilon(p_0)p\sqrt{1-x}$, and $S^{(2)}(q,p;p_0)$ the distributions,
$$S^{(2)}_a(q,p;p_0)={1\over 2}\ {\rm{disc}}_{P'}\int_0^{x'_M(x)}\!\! {{\rm
d}x'\over (P'^2)_R^{N'+1}}\int_{{\widehat K},{\widehat K'}}\ \ {\widehat
K}\!\cdot\!{\widehat
K'} \  { Tr\left(\rlap / \!P\bigl\lbrace \rlap /\!\Sigma_R\rlap /\!
P\bigr\rbrace\ {{\rlap / \!\widehat K}}\ \rlap / \!P'\left(\rlap
/\!\Sigma'_R\rlap
/\! P'\right)^{N'} {{\rlap / \!\widehat
K'}}\right)\over ({\widehat K}\!\cdot\! P)_R\ .\ .\ ({\widehat
K'}\!\cdot\! P')_R}\eqno(5.27)$$
$$S^{(2)}_b(q,p;p_0)={1\over 2}\ {\rm{disc}}_{P'}\int_0^{x'_M(x)}\!\! {{\rm
d}x'\over (P'^2)_R^{N'+1}}\int_{{\widehat K},{\widehat K'}}\ \ {\widehat
K}\!\cdot\!{\widehat
K'} \  { Tr\left(\rlap / \!P\bigl\lbrace I\!\!I_4\bigr\rbrace\ {{\rlap / \!\widehat K}}\ \rlap / \!P'\left(\rlap
/\!\Sigma'_R\rlap
/\! P'\right)^{N'} {{\rlap / \!\widehat
K'}}\right)\over ({\widehat K}\!\cdot\! P)_R\ .\ .\ ({\widehat
K'}\!\cdot\! P')_R}\eqno(5.28)$$In the $R/A$ real time formalism we are using, the imaginary part of
$(N,N';2)$ is effectively obtained out of (5.1), by forming the difference (divided by a factor of 2) of
{\it{retarded}} and {\it{advanced}} $P'$-lines. Whereof results
(after integrating on $x'$) functions of $P$ which exhibit the features
of
distributions rather than of ordinary functions. As displayed for example by (3.43) in the $(N,N';0)$ case, the
discontinuities in $p_0$ of the
$S^{(2)}(q,p;p_0)$ may develop imaginary parts, and this is why they appear inside the discontinuity prescription
of (5.26), and not
simply factored out, as would be
overall real valued multiplicative functions.\smallskip\noindent The
connection with
patterns (5.23) and (5.24) is made complete by recalling that, in virtue of (3.5) and (3.6), the coefficients
$a_{N}$ and $b_{N}$
are polynomials of degree $j_M(N)$ in the variable $(-P^2\Sigma_R^2/m^4)$. We have
then, for all $n$, $0\leq n\leq j_M(N)$,
$$\int_0^1{\rm{d}}x\ {\rm{disc}}_{P}\left({(-P^2\Sigma_R^2)^n\ S^{(2)}\over
(P^2)_R^{N+1}}\right)=\sum_{l=0}^\infty
{1\over
l!}\sum_{i=0}^nC_n^i(-1)^i({1\over 4})^{n-i}\sum_{k=0}^iC_i^k(-1)^k\times (5.23)\eqno(5.29)$$
\medskip

 Now, the statement (5.23)-(5.24) is rather obvious indeed, because expression (5.25) is nothing but a linear
 combination  of  terms  whose general form reads $$x^a{\sqrt{1-x}}^b\ [\epsilon(p_0)\ln X]^c\ ,\ \ \ \
0\leq a,b\leq 1\ ,\ \ \ \ 0\leq
c\leq 2\eqno(5.30)$$The
first contribution of
(5.25) to (5.24), is at $a=b=c=0$, and up to an overall multiplicative factor of $-m^4/p^2$, leaves (5.23) the
same
as it is. The second contribution, is at $a=1,b=0,c=2$. Up to an overall multiplicative factor of $m^4/4p^2$, this
contribution leaves (5.23) unchanged, but for the only modification brought about by the shift of integer number $n-i$, 
 $$(n-i)\ \longmapsto\  (n-i)+1\eqno(5.31)$$The third contribution is at $a=0,b=1,c=1$, and up to an overall
 multiplicative factor of $m^4/p^2$ , it is
 entirely contained in the shift of integer number $k$, with
 $$k\ \longmapsto\  k+1\eqno(5.32)$$It results that, if (5.23) has an {\it{msf}} imaginary part, then, so does
 (5.24). Somehow conversely, the very structure of mass singularity compensation patterns (5.23), makes it
clear
 that if (5.29) is {\it{msf}}, then so is the case of the same whole expression, but taken at $n-1$
 instead  of  $n$.   \bigskip    Next  we  need  to  proof  that  if  (5.22)  has  an  {\it{msf}}  imaginary 
part,  then,  so
 does  
 $$\displaylines{-8ie^2m^4\int {p^2{\rm d}p\over
(2\pi)^2}\sum_{\epsilon(p_0)}\int_0^1 {{\rm d}x\over
2\pi}\   {p_0(x)\over 2T}\ {\rm{disc}}\  {1\over (P^2)_R^{N+2}}\cr\hfill\times\int_0^{x'_M}\!\! {{\rm
d}x'\over (P'^2)_R^{N'+1}}\int_{{\widehat K},{\widehat K'}}\ \ {\widehat
K}\!\cdot\!{\widehat
K'}\   { Tr\left(..\left(\rlap /\!\Sigma_R\rlap /\!
P\right)^{N}\ ..\right)\over ({\widehat K}\!\cdot\! P)_R\ . .\ ({\widehat
K'}\!\cdot\! P')_R}\qquad(5.33)}$$where the difference with (5.22) is that, inside the trace, we have now a power
of $N$
instead of $N+1$. This statement is again rather obvious if one considers that the only change is in the
substitution of the couple of polynomials $(a_{N+1},b_{N+1})$ by the couple $(a_{N},b_{N})$. However, the 
proof of the above statement can be obtained by induction either, assuming first that the involvement
holds true at $(N+1,N';2)$, that is between (5.22) and (5.33). At next order, $(N+2,N';2)$ is given by (5.16),
which decomposes into the sum (5.19). Now, it has just been assumed that (5.22), and hence (5.33) have
{\it{msf}} imaginary parts. In view of statement (5.23)-(5.24), the same is therefore true of the second term in
(5.19), which only
differs (5.33) a multiplicative function of $\Sigma_R^2$. It results that if $(N+2,N';2)$ has an {\it{msf}}
imaginary part, then
so
 does have the expression
 $$\displaylines{ 8ie^2m^4\int {p^2{\rm d}p\over
(2\pi)^2}\sum_{\epsilon(p_0)}\int_0^1 {{\rm d}x\over
2\pi}\   {p_0(x)\over 2T}\  {\rm{disc}} {1\over (P^2)_R^{N+3}}\cr\hfill\times\int_0^{x'_M} {{\rm
d}x'\over (P'^2)_R^{N'+1}}\int_{{\widehat K},{\widehat
K'}}\ \ {\widehat
K}\!\cdot\!{\widehat
K'}\    { Tr\left(\rlap / \!P\left(\rlap /\!\Sigma_R\rlap /\!
P\right)^{N+1}\ {{\rlap / \!\widehat K}}\ \rlap / \!P'\left(\rlap
/\!\Sigma'_R\rlap
/\! P'\right)^{N'} {{\rlap / \!\widehat
K'}}\right)\over ({\widehat K}\!\cdot\! P)_R({\widehat
K}\!\cdot\! P')_R({\widehat K'}\!\cdot\! P)_R({\widehat
K'}\!\cdot\! P')_R}\qquad(5.34)}$$That is, the involvement extends from $(N+1,N';2)$ to $(N+2,N';2)$. Eventually,
we learn out of Ref.13, that $(0,0;2)$ and $(1,0;2)$ have {\it{msf}} imaginary parts. Whereof it is immediate to
check that the property of involvement under consideration is verified at $N=1$; and thus, at all $N$.\bigskip
Getting back to our central induction hypothesis that $(N+1,N';2)$ has an {\it{msf}} imaginary part, the above
two statements allow
to conclude that (5.21) does have an {\it{msf}} imaginary part either, and this establishes that the imaginary
part
of
the difference (5.20) is {\it{msf}}.\bigskip Two possibilities have to be considered whereupon: Either mass
singular behaviours of both members compensate for each others in
the
difference (5.20), or both members of (5.20) have, separately, {\it{msf}} imaginary parts.\smallskip
Let us suppose that a compensation of singularities is at the origin of the difference {\it{msf}} imaginary part.
The trace of (5.20) can be written as
$$Tr\left(\rlap /\!
P\ \left(\rlap /\!\Sigma_R\rlap /\!
P\right)^{N+1}\Bigl\lbrace\rlap /\!\Sigma_R\rlap /\!
P-2m^2\Bigr\rbrace\ {{\rlap / \!\widehat K}}\ \rlap / \!P'\left(\rlap
/\!\Sigma'_R\rlap
/\! P'\right)^{N'} {{\rlap / \!\widehat
K'}}\right)\eqno(5.35)$$As (5.35) stands, however, inspection shows that nothing conclusive can be derived.
Relying again on (3.4)-(3.6),
it is interesting to decompose (5.35) into a sum of terms $$\displaylines{(\Delta a) a_{N'}Tr\left(\rlap /\!
P\left(\rlap /\!\Sigma_R\rlap /\!
P\right){{\rlap / \!\widehat K}}\ \rlap / \!P'\left(\rlap
/\!\Sigma'_R\rlap
/\! P'\right) {{\rlap / \!\widehat
K'}}\right)   +(\Delta a) b_{N'}Tr\left(\rlap /\!
P\left(\rlap /\!\Sigma_R\rlap /\!
P\right){{\rlap / \!\widehat K}}\ \rlap / \!P' {{\rlap / \!\widehat
K'}}\right)\cr\hfill
+(\Delta b) a_{N'}Tr\left(\rlap /\!
P{{\rlap / \!\widehat K}}\ \rlap / \!P'\left(\rlap /\!\Sigma'_R\rlap /\!
P'\right) {{\rlap / \!\widehat
K'}}\right) +(\Delta b) b_{N'}Tr\left(\rlap /\!
P{{\rlap / \!\widehat K}}\ \rlap / \!P'{{\rlap / \!\widehat
K'}}\right)\qquad(5.36)}$$where we have defined
$$\Delta a=a_{N+2}-2m^2a_{N+1}\ ,\ \ \ \Delta b=b_{N+2}-2m^2b_{N+1}\eqno(5.37)$$At its turn, the first trace of
(5.36), decomposes into
a sum of terms,
$$Tr\left(\rlap /\!
P\left(\rlap /\!\Sigma_R\rlap /\!
P\right){{\rlap / \!\widehat K}}\ \rlap / \!P'\left(\rlap
/\!\Sigma'_R\rlap
/\! P'\right) {{\rlap / \!\widehat
K'}}\right)=m^2Tr\left(\rlap /\!
P\left(\rlap /\!\Sigma_R\rlap /\!
P\right){{\rlap / \!\widehat K}}\ \rlap / \!P' {{\rlap / \!\widehat
K'}}\right)+m^2Tr\left(\rlap /\!
P{{\rlap / \!\widehat K}}\ \rlap / \!P'\left(\rlap /\!\Sigma'_R\rlap /\!
P'\right) {{\rlap / \!\widehat
K'}}\right)\ +\ ..\eqno(5.38)$$where the two traces of the right hand side are the second and third traces
of (5.36), respectively, whereas the dots stand for terms which belong, in proper, to the trace under
consideration. If the latter induce further singular imaginary parts, the imaginary part {\it{msf}} character of
(5.20) indicates that mass singularity compensations are taking place, that is,
$$\Delta a=0\eqno(5.39)$$If $N$ is an even number, then, $j_M(N+1)=j_M(N+2)=N/2$, and one has
$$\Delta a=(m^2)^{N+1}\ \sum_{j=0}^{N/2}(C_{N+2}^{2j+1}-2C_{N+1}^{2j+1})\sum_{n=0}^j C_j^n(-{P^2\Sigma^2_R\over
m^4})^n\eqno(5.40)$$For some given powers $n$ to induce singular integrations to be further compensated in the
difference, the following
condition must therefore be satisfied,
$$\sum_{j=n}^{N/2}C_j^n(C_{N+2}^{2j+1}-2C_{N+1}^{2j+1})=0\eqno(5.41)$$Binomial coefficients are
positive definite, and if $n>N/4$, then, the terms in the sum (5.41) are positive definite either, precluding any
compensations of possible singular contributions. That is, contributions attached to the range of powers $N/4<
n\leq N/2$ are necessarily {\it{msf}} in imaginary parts, separately. Now, so are also all of the other powers,
$1\leq n\leq N/2$, in virtue of the statement "somehow reciprocal" to (5.24). Since a similar argument can be
developed in
case
of an odd number $N$, it results that the dots of (5.38) induce, in both members of the difference (5.20),
contributions whose
imaginary parts are {\it{msf}}, separately.\medskip We consider the second trace, and note that in view of (5.36)
and (5.38), it has coefficient
$$(b_{N'}+m^2 a_{N'})\Delta a\eqno(5.42)$$If, when plugged into (5.20), the second trace of (5.36)
generates non {\it{msf}} imaginary parts, then the latter have to compensate each others in the difference.
Now, selecting a power of
$n'$ in the variable $(-{P'^2\Sigma'^2_R/
m^4})$, its coefficient reads
$$\sum_{j'=n'}^{N'/2}\ (C_{j'}^{n'-1}C_{N'-1}^{2j'+1}+C_{j'}^{n'}C_{N'}^{2j'+1})\eqno(5.43)$$where an even value
of
$N'$ is choosen, for the sake of illustration. It is clear that for all $n'\in \{1,2,..N'/2\}$, (5.43)
is a
never vanishing quantity. A compensation of possible singular subsequent integrations on $x$, can only come from
(5.39), with the
conclusion that for this second trace of (5.36), both members of the difference (5.20) have, separately,
{\it{msf}}
imaginary parts.\medskip
The third trace of (5.36) comes into play with a coefficient of
$$a_{N'}(\Delta b +m^2\Delta a)\eqno(5.44)$$which may be explicited as
$$\displaylines{a_{N'}(m^2)^{N+2}\biggl\lbrace
\sum_{j=0}^{N/2}(C_{N+2}^{2j+1}-2C_{N+1}^{2j+1})(1-{P^2\Sigma^2_R\over m^4})^j\cr\hfill+(-{P^2\Sigma^2_R\over
m^4})\sum_{j=0}^{N/2-1}(C_{N+1}^{2j+1}-2C_{N}^{2j+1})(1-{P^2\Sigma^2_R\over
m^4})^j +(-{P^2\Sigma^2_R\over
m^4})(1-{P^2\Sigma^2_R\over
m^4})^{N/2}\biggr\rbrace\qquad(5.45)}$$The higher power in the variable
$-{P^2\Sigma^2_R/ m^4}$ is a power of $N/2+1$, with coefficient $1$. There is no available compensation
for this isolated term of (5.45) which has accordingly to yield a regular subsequent integration on $x$. So is
therefore the case of all of the powers $n\in \{1,2,..N/2+1\}$, because of the statement reciprocal to (5.24).
Whereof results that, irrespective of possible compensations among the other (regular) terms of (5.45),
both members of the difference (5.20) have {\it{msf}} imaginary parts attached to the third trace of
(5.36).\medskip
The fourth trace of (5.36) has coefficient,
$$(m^2)^2a_{N'}\Delta a+m^2(a_{N'}\Delta b+b_{N'}\Delta a)+b_{N'}\Delta b=(b_{N'}+m^2 a_{N'})(\Delta b +m^2\Delta
a)\eqno(5.46)$$If this trace generates any singular subsequent integration on $x$, when put back into both
members of (5.20), the {\it{msf}} character of the imaginary part of
(5.20) requires that $\Delta b +m^2\Delta a$ vanishes. This condition turns out to be the one just dealt with,
and
it results that, when put into both members of the difference (5.20), the fourth trace of (5.36) induce
subsequent
$x$-integrations that have, separately, {\it{msf}} imaginary parts.\bigskip
To summarize, both $(0,0;2)$ and $(1,0;2)$ diagrams, have been shown to possess {\it{msf}} imaginary
parts in Ref.13. Then, assuming that a diagram $(N+1,N';2)$ has {\it{msf}} imaginary
part, we have been able to prove that the next diagram
$(N+2,N';2)$, with one more $HTL$-self energy insertion, has an {\it{msf}} imaginary part either. We can
therefore
conclude that
any of the two effective vertex diagrams contribute {\it{msf}} imaginary
parts to the soft real photon emission rate.\bigskip
The power and simplicity of the proof just developed appears the more clearly as one realizes that the
distributions
$S^{(2)}(q,p;p_0)$ introduced in full generality in (5.23), entail the double vertex function $W_2(P,P')$ and
$W_1(P,P')$  defined in (5.7) and (5.15) respectively. We have in effect, for the $S^{(2)}(q,p;p_0)$ the
expressions $$\displaylines{{1\over 2}\ {\rm{disc}}_{P'}\int_0^{x'_M(x)}\!\! {\rm
d}x'\ {a_{N'}(-P'^2\Sigma'^2/m^4)\over (P'^2)_R^{N'+1}}\int_{{\widehat K},{\widehat K'}}\ \ {\widehat
K}\!\cdot\!{\widehat
K'} \  { Tr\left(\rlap / \!P\bigl\lbrace \rlap /\!\Sigma_R\rlap /\!
P,\ I\!\!I_4\bigr\rbrace\ {{\rlap / \!\widehat K}}\ \rlap / \!P'\bigl\lbrace\rlap
/\!\Sigma'_R\rlap
/\! P'\bigr\rbrace {{\rlap / \!\widehat
K'}}\right)\over ({\widehat K}\!\cdot\! P)_R\ .\ .\ ({\widehat
K'}\!\cdot\! P')_R}\cr\hfill {1\over 2}\ {\rm{disc}}_{P'}\int_0^{x'_M(x)}\!\! {\rm
d}x'\ {b_{N'}(-P'^2\Sigma'^2/m^4)\over (P'^2)_R^{N'+1}}\int_{{\widehat K},{\widehat K'}}\ \ {\widehat
K}\!\cdot\!{\widehat
K'} \  { Tr\left(\rlap / \!P\bigl\lbrace \rlap /\!\Sigma_R\rlap /\!
P,\ I\!\!I_4\bigr\rbrace\ {{\rlap / \!\widehat K}}\ \rlap / \!P'\bigl\lbrace I\!\!I_4\bigr\rbrace {{\rlap /
\!\widehat
K'}}\right)\over ({\widehat K}\!\cdot\! P)_R\ .\ .\ ({\widehat K'}\!\cdot\! P')_R} \qquad(5.47)}$$and whereas the
second term entails $W_2(P,P')$, the first one
entails both $W_2(P,P')$ and $W_1(P,P')$ which are so complicated functions of $P,P'$, that they practically exclude
any control of the ensuing integrations on $x'$, and {\it{a posteriori}} on $x$, contrarily to what could be achieved in Sections 3 and 4,
for
the topologies $(N,N';0)$ and $(N,N';1)$.\medskip An interesting byproduct of this analysis is obtained by
writing, $$S^1_a(q,p;p_0)={1\over 2}\ {\rm{disc}}_{P'}\int_0^{x'_M(x)}\!\! {{\rm
d}x'\over (P'^2)_R^{N'+1}}\int{{\rm d}{\widehat K}\over 4\pi}\  { Tr\left(\rlap / \!P\bigl\lbrace \rlap
/\!\Sigma_R\rlap /\!
P\bigr\rbrace\ {{\rlap / \!\widehat K}}\ \rlap / \!P'\left(\rlap
/\!\Sigma'_R\rlap
/\! P'\right)^{N'} {{\rlap / \!\widehat
K}}\right)\over ({\widehat K}\!\cdot\! P)_R({\widehat
K}\!\cdot\! P')_R}\eqno(5.48)$$
$$S^1_b(q,p;p_0)={1\over 2}\ {\rm{disc}}_{P'}\int_0^{x'_M(x)}\!\! {{\rm
d}x'\over (P'^2)_R^{N'+1}}\int{{\rm d}{\widehat K}\over 4\pi}\  { Tr\left(\rlap / \!P\bigl\lbrace
I\!\!I_4\bigr\rbrace\ {{\rlap / \!\widehat K}}\ \rlap / \!P'\left(\rlap
/\!\Sigma'_R\rlap
/\! P'\right)^{N'} {{\rlap / \!\widehat
K}}\right)\over ({\widehat K}\!\cdot\! P)_R({\widehat
K}\!\cdot\! P')_R}\eqno(5.49)$$and by recognizing that the imaginary part of a diagram $(N,N';1)$ is hereby
expressed
as,
$$8e^2m^2\int {p^2{\rm d}p\over
(2\pi)^2}\sum_{\epsilon(p_0)=\pm 1}\int_0^1 {{\rm d}x\over
2\pi}\   {p_0(x)\over 2T}\ {\rm{disc}}_P\ {a_NS^1_a+b_NS^1_b\over
(P^2)_R^{N+1}}\eqno(5.50)$$Then, knowing from Ref.13 that $(0,0;1)$, $(1,0;1)$ and $(1,1;1)$ have {\it{msf}}
imaginary parts, the same steps as
followed throughout this section can be taken, and allow to conclude by induction that $(N,N';1)$ diagrams
contribute {\it{msf}}
parts to the soft photon emission rate. This is the more systematic derivation which was advertised
in
the end of Sec.4 : It encompasses all of the terms (4.5)-(4.7) of the $(N,N';1)$ situation, and not solely
the
peculiar one, (4.8), which was treated there.\medskip Likewise, identifying now $S^0_a(q,p;p_0)$ and $S^0_b
(q,p;p_0)$ the distributions,
$$S^0_a(q,p;p_0)={1\over 2}\ {\rm{disc}}_{P'}\int_0^{x'_M(x)}\!\! {{\rm
d}x'\over (P'^2)_R^{N'+1}}\  { Tr\left(\rlap / \!P\bigl\lbrace \rlap
/\!\Sigma_R\rlap /\!
P\bigr\rbrace\ \rlap / \!P'\left(\rlap
/\!\Sigma'_R\rlap
/\! P'\right)^{N'} \right) }$$
$$S^0_b(q,p;p_0)={1\over 2}\ {\rm{disc}}_{P'}\int_0^{x'_M(x)}\!\! {{\rm
d}x'\over (P'^2)_R^{N'+1}}  { Tr\left(\rlap / \!P\bigl\lbrace
I\!\!I_4\bigr\rbrace\ \rlap / \!P'\left(\rlap
/\!\Sigma'_R\rlap
/\! P'\right)^{N'} \right)}$$and observing that the imaginary part of a diagram $(N,N';0)$ can be expressed as
$$8e^2\int {p^2{\rm d}p\over
(2\pi)^2}\sum_{\epsilon(p_0)=\pm 1}\int_0^1 {{\rm d}x\over
2\pi}\   {p_0(x)\over 2T}\ {\rm{disc}}_P\ {a_NS^0_a+b_NS^0_b\over
(P^2)_R^{N+1}}$$
we can make use of the $msf$ character of $(1,0;0)$ and $(2,0;0)$ imaginary parts as established in Ref.13 so as
to
follow the
same steps as taken throughout this section and conclude, in agreement with the
calculational approach of Sec.3, that $(N,N';0)$ diagrams contribute $msf$ parts to the soft photon emission
rate.

\bigskip\bigskip\bigskip\bigskip\bigskip\bigskip\bigskip\bigskip\bigskip\bigskip\bigskip\bigskip\bigskip\bigskip
{\bf{ VI. Conclusion}}
\bigskip
Some years ago$^{23}$, we had suggested that the collinear problem met in hot $QCD$ when the {\it{Resummation
Program}} ($RP$) is used, could be traced back to the particular perturbative series re-arrangement the $RP$
amounts to. Strictly speaking though (and contrarily to what can be read off the existing
literature) the $RP$ should not be mistaken for any Feynman diagrams resummation,
possibly infinite. This happens
to be so, simply
because of the effective propagators non perturbative character : Pole residues and dispersion relations, in
effect,
cannot
be derived out of pure thermal Perturbation Theory.\smallskip This suggestion has motivated our
construction of a coherent {\it{Perturbative Resummation}} scheme ($P\!R$) of the leading thermal effects (the
so-called {\it{Hard Thermal Loops}}, $HTL$) enjoying by construction the same symmetry properties as the usual
$RP$, with the hope that things could come out at variance with the troublesome (undefined!) $RP$ results.
\smallskip In the case of the so called {\it{rapid fermion damping rate}} problem of both $QED$ and
$QCD$, a first obstruction met by the $RP$$^5$, this hope revealed itself non deceptive indeed$^{6,7}$, whereas
the
collinear problem under consideration was subsequently projected out on a simpler toy model, with promising
results$^{11}$.\smallskip In a recent publication$^{13}$, the physically interesting case of hot $QCD$
has been analyzed through its first non trivial perturbative orders, with very instructive new insights. As stated
in
the introduction, not only did the $P\!R$ analysis allowed to elucidate the so far questionable nature$^{12}$ of
the collinear singularity encountered in hot $QCD$, when the
$RP$
is used, but a tight and original comparison of both $RP$ and $P\!R$ calculations made it
possible to understand how the collinear singularity unavoidably shows up in an $RP$ treatment.\medskip

Now, a $P\!R$ calculation of the soft real photon emission rate, involves the infinite
resummations on $N$ and $N'$, of any perturbative contribution of type $(N,N';0)$, $(N,N';1)$ and $(N,N';2)$,
describing one loop photonic self energy diagrams at
$N$($N'$) $HTL$ self energy insertions along the $P$($P'$) -fermionic line, respectively endowed with zero, one
and
two $HTL$ effective vertex corrections. In order to set our $P\!R$ calculation a sound, significant result, in
contradistinction to
the yet confused $RP$ situation, it was therefore crucial to check that the properties that could be derived for
the
perturbative orders of
$m^2$, $m^4$ and $m^6$, extended indeed to any contribution of order $m^{2n}$, and this is the task which has been
achieved throughout the present article. \medskip
However tedious the calculational developments of Section 3, they eventually revealed extremely useful so as to
provide a proof by induction with sound enough a basis. In other words, a proof by induction is not reliable
until enough information is gained concerning the mass singularity cancellation patterns, and not before.

Physically, a salient aspect comes out to be the one of propagator's pole
migration. This migration in effect, from the light-like to the
time-like region, appears to be at the very origin of the $RP$ dramatic consequence
under
consideration, and in contrast to the $T=0$ situation, is really peculiar to the thermal context$^{7}$.
This is because pole displacements involve a decoupling of partial effective
propagators
(potential) mass singularities, from effective vertices collinear ones : Whereas all singularities mix
up
into patterns which
grant their overall compensations in a $P\!R$ calculation, effective vertices mass/collinear singularities
remain isolated in an $RP$ calculation, with no singular counterpart to cancel against. This mechanism, first
guessed in Ref.11, then discovered in Ref.13 for the perturbative orders of $m^2$, $m^4$ and $m^6$, is easily
seen here to spread to any
perturbative orders, $(m^2)^n$. \medskip

As we have long been suspecting$^{23}$, the collinear singularity plaguing the soft real photon emission rate $RP$
calculation is likely to be nothing but an {\it{artefact}},
peculiar to the $RP$ resummation scheme itself. Here may be the place where to recall a part of our conclusion in
Ref.13: {\it{After all, whenever resummation is required by the context, a guiding principle could very well be
that
it be conceived and taken out of finite, well defined elements, and in particular, of mass singularity free
terms.
In this respect, it is instructive to come back to the original article where the $RP$ was mostly founded, and to
realize that the authors were conscious of difficulties that could be inherited from the fact that the $RP$ did
not
necessarily comply with this requirement$^{24}$.}}
\bigskip\bigskip\bigskip
{\bf {Acknowledgement}} \medskip It is a pleasure to thank B.Candelpergher for encouraging discussions.
\bigskip\bigskip\bigskip

{\bf{Appendix: Calculating the $W_i(P,P')$}}
\bigskip
In this appendix account is given of the difficulty inherent to the explicit calculations of the double effective
vertex functions $W_i(P,P')$, $i=1,2$. The first function, $W_1(P,P')$ is given in (5.15). It is given by the
integration on a Feynman parameter $s$, of the squared norm of a {\it{would be}}
$4$-vector with components the right hand sides of (4.9) and (4.10),
$$\displaylines{m^2\int{{\rm d}{\widehat K}\over 4\pi}{{\widehat
K}^\mu\over ({\widehat K}\!\cdot\! P+i\epsilon)({\widehat
K}\!\cdot\! P'+i\epsilon)}\ m^2\int{{\rm d}{\widehat K'}\over 4\pi}{{\widehat
K'}_\mu\over ({\widehat K'}\!\cdot\! P+i\epsilon)({\widehat
K'}\!\cdot\! P'+i\epsilon)}=\cr\hfill m^4\int_0^1{\rm{d}}s\left({r_0\over r^3}{\ln X_r\over R^2+i\epsilon
r_0}-{1\over 4r^4}\ln^2 X_r -{1\over r^2(R^2+i\epsilon r_0)}\right)\qquad(A.1)}$$As it stands however, the
remaining integration on $s$ is not very easy. A more economic way to proceed consists in
writing $$\displaylines{m^4\int{{\rm d}{\widehat K}\over 4\pi}{{\widehat
K}^\mu\over ({\widehat K}\!\cdot\! P+i\epsilon)({\widehat
K}\!\cdot\! P'+i\epsilon)}\ \int{{\rm d}{\widehat K'}\over 4\pi}{{\widehat
K'}_\mu\over ({\widehat K'}\!\cdot\! P+i\epsilon)({\widehat
K'}\!\cdot\! P'+i\epsilon)}=\cr\hfill\int_0^1{\rm{d}}s\int_0^1{\rm{d}}s'\ m^2\int{{\rm d}{\widehat K}\over
4\pi}{{\widehat K}^\mu\over ({\widehat K}\!\cdot\! R(s)+i\epsilon)^2}\ m^2\int{{\rm d}{\widehat K'}\over
4\pi}{{\widehat K'}_\mu\over ({\widehat K'}\!\cdot\! R(s')+i\epsilon')^2}\qquad(A.2)}$$This allows to write (A.1)
as,
$$\displaylines{(-{{\rm{d}}\over {\rm{d}} i\epsilon})(-{{\rm{d}}\over {\rm{d}} i\epsilon'})\
\int_0^1{\rm{d}}s\int_0^1{\rm{d}}s'\ \Sigma\left(R(s)\right)\!\cdot\!\Sigma\left(R(s')\right)\cr\hfill
=\int_0^1{\rm{d}}s\int_0^1{\rm{d}}s'\ {m^2\over R^2(s)+i\epsilon r_0(s)}{m^2\over R^2(s')+i\epsilon r_0(s')}\
\left(1-{r_0\over r}(s){r_0\over r}(s')\ {\widehat
r}\!\cdot\!{\widehat
r'}\right)\qquad(A.3)}$$The term $+1$ in the right hand side last parenthesis yields simply,
$$\int_0^1{\rm{d}}s\int_0^1{\rm{d}}s'\ {m^2\over R^2(s)+i\epsilon r_0(s)}{m^2\over R^2(s')+i\epsilon
r_0(s')}= \left({m^2\over 2P\!\cdot Q}\ln{P'^2\over P^2}\right)^2\eqno(A.4)$$The contribution of the term
involving the cosine ${\widehat
r}\!\cdot\!{\widehat
r'}$ is of course more involved. It is
$$\displaylines{-\int_0^1{\rm{d}}s\ {m^2\over R^2(s)+i\epsilon r_0(s)}\ {p_0+qs\over r^2(s)}\cr\hfill
\times\ 
\int_0^1{\rm{d}}s'\ {m^2\over R^2(s')+i\epsilon r_0(s')}\ {p_0+qs'\over
r^2(s')}\ \left(p(p+qys)+q(py+qs)s'\right)\qquad(A.5)}$$where,
$$r^2(s)=p^2+2pqys+q^2s^2\ ,\ \ \ \ R^2(s)=P^2+2P\!\cdot Qs\eqno(A.6)$$Introducing the three
functions,
$$F_1(P,Q)={1\over 2Q\!\cdot\! P}\ln{P'^2\over P^2}\eqno(A.7)$$
$$F_2(P,Q)={1\over qp{\sqrt{1-y^2}}}\arctan{q{\sqrt{1-y^2}}\over p+qy}\eqno(A.8)$$
$$\displaylines{F_3(P,Q)={1\over (P^2{\vec{p'}}-{\vec{p}}P'^2)^2}\Bigl\lbrace{1\over 2}(2Q\!\cdot\!
P)^2F_1(P,Q)\cr\hfill -(q^2P^2-qp_0 (2Q\!\cdot\! P)+{1\over 2}(2Q\!\cdot\! P)^2
)F_2(P,Q)\Bigr\rbrace\qquad(A.9)}$$one verifies that (A.4) cancels out, and that the integration of (A.1) can be
given the interesting following form, $$\displaylines{m^2\int{{\rm d}{\widehat K}\over
4\pi}{{\widehat
K}^\mu\over ({\widehat K}\!\cdot\! P+i\epsilon)({\widehat
K}\!\cdot\! P'+i\epsilon)}\ m^2\int{{\rm d}{\widehat K'}\over 4\pi}{{\widehat
K'}_\mu\over ({\widehat K'}\!\cdot\! P+i\epsilon)({\widehat
K'}\!\cdot\! P'+i\epsilon)}=\cr\hfill -m^4\sum_{i,j=1}^3 \ \left(\sum_{k=-2}^{+1} a^k_{ij}\ (2Q\!\cdot\!
P)^k\right)
F_iF_j\qquad(A.10)}$$where the non vanishing $a^k_{ij}$ coefficients are given by

$$a^{-2}_{22}=-q^2P^2,\  a^{-1}_{22}=qp_0\eqno(A.11)$$
$$a^{-2}_{33}=-q^2(P^2)^3,\ a^{-1}_{33}={5\over 2}qp_0(P^2)^2,\ a^{0}_{33}=-{9\over 4}(P^2)^2-{5\over 2}p^2P^2,\
a^1_{33}={p_0(3P^2+4p^2)\over 4q}\eqno(A.12)$$
$$a_{12}^0=1\eqno(A.13)$$
$$a_{13}^{-1}=-qp_0P^2,\ a_{13}^0={3\over 2}P^2,\ a_{13}^1=-{p_0\over q}\eqno(A.14)$$
$$a_{23}^{-2}=2q^2(P^2)^2,\ a_{23}^{-1}=-4qp_0P^2,\ a_{23}^0={11\over 4}P^2+{3\over
2}p^2,\ a_{23}^1=-{p_0\over q}\eqno(A.15)$$
\bigskip\bigskip
Calculating $W_2(P,P')$, given in (5.7), is the most tedious angular integration to be coped
with, "an order of magnitude" more difficult than the latter. One has,
$$W_2(P,P')=\int_0^1{\rm{d}}s\int_0^1{\rm{d}}s'\int{{\rm d}{\widehat K}\over
4\pi}\int{{\rm d}{\widehat K'}\over
4\pi}{1-2{\widehat K}^i{\widehat K'}_i+{\widehat K}^i{\widehat K}^j{\widehat K'}_i{\widehat K'}_j\over ({\widehat
K}\!\cdot\!R(s)+i\epsilon)^2({\widehat K'}\!\cdot\!R(s')+i\epsilon)^2}\eqno(A.16)$$The full set of angular
identities (4.9)-(4.11) must be used, but since the first two terms in the numerator of (A.16) are those which
have just been dealt with in the calculation of $W_1(P,P')$, we may focus on the contribution due to the third
term, $\{{\widehat K}^i{\widehat K}^j{\widehat K'}_i{\widehat K'}_j \}$. Relying on the angular identity (4.11),
one finds, with $Q_1$ as defined in (3.3), $$\displaylines{-3\left( \int_0^1{\rm{d}}s\ {Q_1(R(s))\over
r^2(s)}\right)^2+6\left( \int_0^1{\rm{d}}s\ {Q_1(R(s))\over
r^2(s)}\right)\left( \int_0^1\ {{\rm{d}}s'\over
R^2(s')+i\epsilon r_0(s')}\right)\cr\hfill +9\int_0^1{\rm{d}}s\ {Q_1(R(s))\over
r^2(s)}\int_0^1{\rm{d}}s'\ [{\widehat r(s)}\!\cdot\!{\widehat r(s')}]^2{Q_1(R(s'))\over
r^2(s')}\ -6\int_0^1{\rm{d}}s\ {Q_1(R(s))\over
r^2(s)}\int_0^1{\rm{d}}s'\ {[{\widehat r(s)}\!\cdot\!{\widehat r(s')}]^2\over
R^2(s')+i\epsilon r_0(s')}\cr\hfill +\int_0^1{{\rm{d}}s\over (R^2(s)+i\epsilon r_0(s))}\int_0^1{\rm{d}}s'\
{[{\widehat r(s)}\!\cdot\!{\widehat r(s')}]^2\over (R^2(s')+i\epsilon r_0(s'))}\qquad(A.17)}$$The second term
gives
$$6F_1\int_0^1{\rm{d}}s\ {Q_1(R(s))\over r^2(s)}={3F_1\over pq(1-y^2)}\left((p_0y-p+{Z\over 2p}){\ln X'\over
p'}-(p_0y-p){\ln X\over p}\right)-{3\over 2}{ (ZF_1)^2\over p^2q^2(1-y^2)}\eqno(A.18)$$where the shorthand
notation
$Z=2Q\!\cdot\!P$ has been introduced. The first term of (A.17) gives
$$\displaylines{\!\!\!\!\!\!\!\!\!\!\!\!\!\!\!\!\!\!\!\!\! \!\!\!\!\!\!\!\!\!\!\!\!\!\!\!\!\!\!\!\!\!
\!\!\!\!\!\!\!\!\!\!\!\!\!\!\!\!\!\!\!\!\! \!\!\!\!\!\!\!\!\!\!\!\!\!\!\!\!\!\!\!\!\!
\!\!\!\!\!\!\!\!\!\!\!\!\!\!\!\!\!\!\!\!\! \!\!\!\!\!\!\!\!\!\!\!\!\!\!\!\!\!\!\!\!\! -3\left( \int_0^1{\rm{d}}s\
{Q_1(R(s))\over
r^2(s)}\right)^2\cr\hfill ={-3\over 4p^2q^2(1-y^2)^2}\left(-{1\over 2}{Z^2F_1\over pq}+(p_0y-p+{Z\over 2p}){\ln
X'\over p'}-(p_0y-p){\ln X\over p}\right)^2\qquad(A.19)}$$The fifth term of (A.17) gives
$$\sum_{i,j=1}^3 \ \left(\sum_{k=-2}^{+1} c^k_{ij}\ Z^k\right)
F_iF_j\eqno(A.20)$$where the non vanishing $c^k_{ij}$ are given by,
$$c^0_{11}=1\eqno(A.21)$$
$$c^{-2}_{22}=2{q^2p^2(1-y^2)}\eqno(A.22)$$
$$c^{-2}_{33}=-2{q^2p^4(1-y^2)^2P^2}\ ,\ \ \ c_{33}^{-1}=-2qp_0p^2(1-y^2)P^2\ ,\ \ \ c_{33}^0={p^2+3p_0^2\over
2}\eqno(A.23)$$
$$c_{13}^0=-2p^2(1-y^2)\eqno(A.24)$$
$$c_{23}^{-2}=2{q^2p^2(1-y^2)(-P^2+p^2(1-y^2))}\ ,\ \ \ c_{23}^{-1}=2qp_0p^2(1-y^2)\ ,\ \ \
c_{23}^0=-{3p^2(1-y^2)\over 2}\eqno(A.25)$$The fourth term of (A.17) can first be expressed as,
$$\displaylines{-6\left(F_1-p^2(1-y^2)F_3\right)\left(\int_0^1{\rm{d}}s\ {Q_1(R(s))\over
r^2(s)}\right)+6p^2(1-y^2)\left(F_1-2p^2F_3\right)\left(\int_0^1{\rm{d}}s\ {Q_1(R(s))\over
r^4(s)}\right)\cr\hfill -12qp^3y(1-y^2)F_3\left(\int_0^1{\rm{d}}s\ s{Q_1(R(s))\over
r^4(s)}\right)\qquad(A.26)}$$allowing to see that the second term contribution, (A.18), cancels out with an
identical part
in (A.26). In terms of more elementary integrals, the remaining parts of (A.26) may be written,
$$\displaylines{6F_1F_2-6F_1p^2(1-y^2)\int_0^1{{\rm{d}}s\over r^4(s)}-6F_1\left({1\over 2}\int_0^1{{\rm{d}}s\over
r^3(s)}r_0\ln X_{R(s)}\right)\cr\hfill +6p^2(1-y^2)(F_1-F_3)\left({1\over 2}\int_0^1{{\rm{d}}s\over
r^5(s)}r_0\ln X_{R(s)}\right) \cr\hfill -6F_3p^2(1-y^2)\left(F_2-2p^2\int_0^1{{\rm{d}}s\over
r^4(s)}-2qpy\int_0^1{\rm{d}}s{s\over r^4(s)}\right)
\cr\hfill +12q(1-y^2){F_2-P^2F_3\over Z}\left(p^3y\int_0^1{{\rm{d}}s\over
r^4(s)}+qP^2\int_0^1{\rm{d}}s{s\over
r^4(s)}\right)\cr\hfill -6qp^2(1-y^2){F_2-P^2F_3\over Z}\biggl\lbrace\int_0^1{\rm{d}}s{\ln X_{R(s)}\over
r^3}+p(yp_0-p)\int_0^1{\rm{d}}s{\ln X_{R(s)}\over
r^5}\cr\hfill +{Z\over 2}\int_0^1{\rm{d}}s{s\ln X_{R(s)}\over
r^5}\biggr\rbrace\qquad(A.27)}$$The third term of (A.17) is the more involved one, and may be written as,
$$\displaylines{9\left( \int_0^1{\rm{d}}s\
{Q_1(R(s))\over
r^2(s)}\right)^2-18p^2(1-y^2)\left( \int_0^1{\rm{d}}s\
{Q_1(R(s))\over
r^2(s)}\right)\left( \int_0^1{\rm{d}}s\
{Q_1(R(s))\over
r^4(s)}\right)\cr\hfill +18p^4(1-y^2)\left( \int_0^1{\rm{d}}s\
{Q_1(R(s))\over
r^4(s)}\right)^2 +36qp^3y(1-y^2)\left( \int_0^1{\rm{d}}s\
{Q_1(R(s))\over
r^4(s)}\right)\left( \int_0^1{\rm{d}}s\
s{Q_1(R(s))\over
r^4(s)}\right)\cr\hfill +18q^2p^2(1-y^2)\left( \int_0^1{\rm{d}}s\
s{Q_1(R(s))\over
r^4(s)}\right)^2\qquad(A.28)}$$To summarize, the last piece of $W_2(P,P')$ can eventually be expressed in the
form,
$$\displaylines{\sum_{i,j=1}^3 \ \left(\sum_{k=-2}^{+1} c^k_{ij}\ Z^k\right)
F_iF_j+6p^2(1-y^2)\biggl\lbrace F_3\left(\int_0^1{\rm{d}}s\ {Q_1\over
r^2}\right)\cr\hfill +\left(F_1-2p^2F_3\right)\left(\int_0^1{\rm{d}}s\ {Q_1\over
r^4}\right)\biggr\rbrace\cr\hfill +12qp^3y(1-y^2)\left(\int_0^1{\rm{d}}s\ s{Q_1\over
r^4}\right)\left(-F_3+3\int_0^1{\rm{d}}s\ {Q_1\over
r^4}\right)\cr\hfill +6\left(\int_0^1{\rm{d}}s\ {Q_1\over
r^2}\right)^2-18p^2(1-y^2)\left(\int_0^1{\rm{d}}s\ {Q_1\over
r^2}\right)\left(\int_0^1{\rm{d}}s\ {Q_1\over
r^4}\right)\cr\hfill +18p^4(1-y^2)\left(\int_0^1{\rm{d}}s\ {Q_1\over
r^4}\right)^2+18q^2p^2(1-y^2)\left(\int_0^1{\rm{d}}s\ s{Q_1\over
r^4}\right)^2\qquad(A.29)}$$\smallskip
We will not proceed further, giving for example the more elementary integrals displayed in (A.27) and (A.29), as
it
should already appear clear that the statement concerning "{\it{so cumbersome calculations that they practically
preclude any peer control of ensuing integrations on $x'$ and then on $x$..}}" is not exaggerated.\medskip
Before concluding this appendix we may stress that the calculation of $W_1(P,P')$ and $W_2(P,P')$ does not
display
singularities other than
(potentially) collinear ones, showing up by the light cone boundary, at $P^2\simeq 0$. For the function $F_3$
of
(A.9), this property may be not so easy to see. However, it is straightforward to check that the denominator of
(A.9) reads as,
$$(P^2{\vec{p'}}-{\vec{p}}P'^2)^2=4q^2p^2p_0^2\ (y-{p_0^2+p^2\over 2p_0p})^2\eqno(A.30)$$It vanishes at the light
cone only, at $p_0=\pm p$, and this corresponds effectively to a collinear singularity, at $y=\pm
1$.

\vfill\eject
{\bf{References}}

\bigskip\bigskip $^{1}$ M. Le Bellac, ``Thermal Field Theory" (Cambridge University
Press, 1996). \smallskip\smallskip\smallskip $^{2}$ N.P. Landsman, "Quark Matter 90",
{\it{Nucl. Phys. A}}
 {\bf{525}}, 397 (1991).\smallskip\smallskip\smallskip $^{3}$ E. Braaten and R. Pisarski,
{\it{Phys. Rev. Lett.}} {\bf{64}}, 1338 (1990).\smallskip {\it{Nucl. Phys. B}}{\bf{337}}, 569
(1990).\smallskip J. Frenkel and J.C Taylor, {\it{Nucl. Phys.}} B{\bf{334}}, 199
(1990).\smallskip\smallskip\smallskip \smallskip $^{4}$ E. Braaten and E. Petitgirard, {\it{Phys. Rev. D}}
{\bf{65}}, 085039
(2002), Sec.4.\smallskip\smallskip\smallskip
$^{5}$ R.D. Pisarski, {\it{Phys. Rev.
Lett.}} {\bf{63
}}, 1129 (1989). \smallskip\smallskip\smallskip $^{6}$ R. Baier, S. Peign\'e and D. Schiff,
{\it{Z. Phys. C}} {\bf{62}}, 337 (1994).\smallskip\smallskip\smallskip  $^{7}$ B. Candelpergher and T. Grandou,
{\it{Ann. Phys. (NY)}} {\bf{283}}, 232 (2000); \smallskip [arXiv:hep-ph/0009349].\smallskip
\smallskip\smallskip $^{8}$ T. Muta, ``Foundations of Quantum Chromodynamics" (World Scientific,
1987).\smallskip\smallskip\smallskip 
$^9$ A. Niegawa, {\it{Mod. Phys. Lett.}} A{\bf{10}}, 379 (1995);\smallskip F. Flechsig and A. Rebhan, {\it{Nucl.
Phys.}}
B{\bf{464}}, 279 (1996).\smallskip\smallskip\smallskip 
$^{10}$ P. Aurenche, F.
Gelis, R. Kobes and E. Petitgirard, {\it{Z. Phys. C}} {\bf{75}}, 315 (1997);\smallskip P. Aurenche, F.
Gelis, R. Kobes and H. Zaraket,\smallskip
{\it{Phys. Rev. D}} {\bf{58}}, 085003 (1998), and references
therein.\smallskip\smallskip\smallskip
$^{11}$ T. Grandou, {\it{Acta Physica Polonica B{\bf{32}}}}, 1185 (2001).\smallskip\smallskip\smallskip
$^{12}$ F. Gelis, Th\`ese pr\'esent\'ee \`a l'Universit\'e de Savoie, le 10 D\'ecembre
1998.\smallskip\smallskip\smallskip
$^{13}$ B. Candelpergher and T. Grandou,
{\it{Nucl. Phys. A}} {\bf{699}}, 887 (2002). 
\smallskip\smallskip\smallskip
$^{14}$ P. Aurenche
and T,
Becherrawy, {\it{Nucl. Phys. B}} {\bf{379}}, 259 (1992);\smallskip M.A. van Eijck and C.G. van Weert, {\it Phys.
Lett. B} {\bf 278}, 305 (1992);\smallskip M.A. van Eijck, R. Kobes and C.G. van Weert, {\it{Phys. Rev. D}}
{\bf{50}}, 4097
(1994).   \smallskip\smallskip\smallskip 
$^{15}$ E. Petitgirard, Th\`ese pr\'esent\'ee \`a l'Universit\'e de Savoie, le 25 F\'evrier
1994\smallskip [arXiv:hep-ph/9403320].
\smallskip\smallskip\smallskip
$^{16}$ E. Braaten, R.D. Pisarski, and T.C. Yuan, {\it{Phys. Rev. Lett.}} {\bf{64}}, 2242
(1990).\smallskip S.M.H. Wong, {\it{Z. Phys. C}} {\bf{53}}, 465 (1992); {\it{Z. Phys. C}} {\bf{58}}, 159 (1993).
\smallskip\smallskip\smallskip $^{17}$ T. Grandou, M. Le Bellac and
D. Poizat, {\it{Nucl. Phys. B}} {\bf{358}}, 408 (1991);\smallskip O. Steinmann, {\it Commun.
Math. Phys.} {\bf 170}, 405 (1995);\smallskip T. Grandou, ``Puzzling aspects of hot quantum fields",\smallskip
Proceedings of
the $5^{th}$ Workshop on $QCD$,\smallskip H.M. Fried, B. M\"uller and Y. Gabellini, Editors,\smallskip World
Scientific 2000[arXiv:
hep-ph/0009351]. \smallskip\smallskip\smallskip
$^{18}$ A.K. Rebhan,
{\it{Phys. Rev. D}} {\bf{46}}, 4779 (1992).\smallskip\smallskip\smallskip
$^{19}$ T. Grandou, in preparation.\smallskip\smallskip\smallskip
$^{20}$ P. Aurenche, E. Petitgirard and T. del Rio Gaztelurrutia,\smallskip Phys. Lett. B
{\bf{297}}, 337 (1992).\smallskip\smallskip\smallskip $^{21}$ A.H. Weldon, Phys.Rev. D{\bf{47}}, 594
(1993).\smallskip\smallskip\smallskip $^{22}$ Some unpublished 1995's preprints.\smallskip\smallskip\smallskip
$^{23}$ T. Grandou,
{\it{Phys. Lett.}} B{\bf{367}}, 229 (1996).\smallskip\smallskip\smallskip
$^{24}$ E. Braaten and R.D. Pisarski, second article of Ref.3, p.p 625-626.
\bigskip\bigskip\bigskip\bigskip\bigskip\bigskip\bigskip\bigskip
{\bf Figure caption}\bigskip\bigskip

{\bf Fig.1:} A graph denoted by $(N,N';0)$, with $N(N')$ insertions of $HTL$ self energy along the
$P(P')$-line, and two bare vertices $-ie\gamma_\mu$ of (2.4). 
\bigskip

{\bf Fig.2:} A graph denoted by $(N,N';1)$, with $N(N')$ insertions of $HTL$ self energy along the
$P(P')$-line, one bare vertex $-ie\gamma_\mu$, and one $HTL$ vertex correction (2.5). 
\bigskip

{\bf Fig.3:} A graph denoted by $(N,N';2)$, with $N(N')$ insertions of $HTL$ self energy along the
$P(P')$-line, and two $HTL$ vertex corrections (2.5).

\end